\documentclass[final,english]{elsarticle}
\usepackage[T1]{fontenc}
\usepackage[latin9]{inputenc}
\usepackage{amsmath}
\usepackage{amssymb}
\usepackage{graphicx}
\usepackage{caption}
\usepackage{subcaption}
\usepackage[separate-uncertainty=true]{siunitx}

\journal{Elsevier}
\usepackage{upgreek}
\usepackage[bookmarks,bookmarksopen,bookmarksdepth=6]{hyperref}
\usepackage{url}

\usepackage{a4wide}

\usepackage{newtxtext,newtxmath}

\usepackage{anyfontsize} 

\usepackage{babel}
\begin{document}


\title{Precision determination of the track-position resolution of beam telescopes}

 \author[]{M.~\texorpdfstring{Antonello\corref{cor1}}{Antonello}}
 \author[]{L.~Eikelmann}
 \author[]{E.~Garutti}
 \author[]{R.~Klanner}
 \author[]{J.~Schwandt}
 \author[]{G.~Steinbr\"uck}
 \author[]{A.~Vauth}

\cortext[cor1]{Corresponding author. Email address: massimiliano.antonello@uni-hamburg.de.}

\address{Institute for Experimental Physics, University of Hamburg,
 \\Luruper Chaussee 149, 22761, Hamburg, Germany.}


\begin{abstract}
Beam tests using tracking telescopes are a standard method for determining the spatial resolution of detectors. This requires the precise knowledge of the position resolution of beam tracks reconstructed at the Device Under Test (DUT). A method is proposed which achieves this using a segmented silicon detector with readout with charge digitization. It is found that the DUT spatial resolution for single minimum ionizing particles with normal incidence is less than \SI{1}{\um} for events where clusters consist of two pixels (or strips). Given this high accuracy, the residuals between the beam track-position at the DUT and the reconstructed position in the DUT represent a direct measurement of the beam track-position resolution distribution. The method is developed using simulated events, which are also used to study how to deal with cross-talk, electronics noise, energetic $\delta $-electrons, and incident beams with a few degrees off the normal to the sensor plane. To validate the method, the position resolution of beam tracks reconstructed by the EUDET beam telescope of the DESY~II~Test~Beam~Facility is determined using a CMS\,Phase-2 prototype pixel sensor.
\end{abstract}


\maketitle
\pagenumbering{arabic}

\section{Introduction}
\label{sect:Introduction}
Beam tests are one of the preferred methods to experimentally determine the spatial resolution of segmented silicon detectors. This is achieved by measuring the residual between the track-position extrapolated from a beam telescope to the Device Under Test (DUT) and the position reconstructed in the DUT. In this paper, the resolution of the extrapolation from the beam telescope of the track-position at the DUT will be also referred to as "track-position resolution". By quadratically subtracting the track-position resolution from the residual, the DUT spatial resolution is obtained. Typically, the DUT spatial resolution is similar to the track-position resolution, therefore a precise knowledge of the latter is important. There are a number of methods for determining the track-position resolution of beam telescopes:
\begin{itemize}
\item Measurement of the residuals at the DUT of the track-positions from two beam telescope arms, one upstream and one downstream of the DUT. Assuming that both arms have the same track-position resolution, scaling the residuals by a factor 1/2 gives the resolution of the average position from the upstream and downstream telescope arms. A problem arises if the resolutions from the upstream and downstream arms at the DUT are different. This is the case if there is significant scattering material downstream of the DUT, which is typically required for the cooling of irradiated silicon sensors.
\item Simulation of the entire setup, which requires the knowledge of the hit resolutions of the individual sensors of the beam telescope, the precise geometry and the thicknesses and radiation lengths of the scattering material in the beam line, potentially leading to systematic errors in the track-position resolution. Additionally, the resolution can also be obtained by parameterised simulations\,\cite{Jansen:2018}.
\item In Refs.\,\cite{Terzo:2015, Koppenhoefer:2022, Ziemons:2022}, events where the cluster in the DUT consists of only one pixel (abbreviated in the following as cluster-size-one events) for a normally incident beam are selected and the residuals of the position of the reconstructed track and the centre of the hit pixel are analysed. If the track-position resolution is small compared to the pixel pitch, a box-shaped distribution with edges smeared by the track-position resolution is observed. By fitting this distribution with the difference of two error functions, the track-position resolution is determined. In the Appendix the main limitations of this method are discussed using experimental data.
\item In Ref.\,\cite{Kwan:2016} the observed DUT resolution of \SI{2.38 \pm 0.60}{\um} for events with clusters composed of two pixels for normally incident particles is used to verify the track-position resolution of the FNAL beam telescope. This validation is achieved by analysing the residuals between the positions reconstructed by the beam telescope and by the DUT. The multiple Coulomb scattering implementation in the track reconstruction algorithm was verified by this analysis.
\end{itemize}

In this paper a method is proposed which, for beam tests of segmented silicon detectors with readout with charge digitization, allows the experimental determination of the track-position resolution of single minimum ionizing particles directly from the data. Similar to Ref.\,\cite{Kwan:2016}, it uses the well-known fact that, for normally incident particles, the spatial resolution of segmented silicon detectors at the boundary between pixels (or strips) is less than \SI{1}{\um}. The method is demonstrated using both simulations and real beam test data, measured at the DESY~II~Test~Beam~Facility\,\cite{Diener:2019}, from a non-irradiated CMS prototype pixel sensor\,\cite{Pixel:2023} for the CMS Phase-2 upgrade\,\cite{CMS-ph2}, read out by the RD53A chip\,\cite{Dimitrievska:2020}. 
In the next section, the method is explained and studied using simulated data. In Section\,\ref{sect:Validation}, a short summary of the DESY~II~Test~Beam~Facility, the pixel sensors used, and the data-taking conditions is provided. The proposed method is then validated using experimental data. In the conclusions, the main results are summarized. Finally, in \ref{app:Comparison}, the method is compared to the one from\,\cite{Terzo:2015, Koppenhoefer:2022, Ziemons:2022} mentioned above.

\section{Method}
\label{sect:Method}
\subsection{Simulation}
\label{subsect:Simulation}
The method described aims to determine the position resolution of tracks reconstructed in a beam telescope and extrapolated to the DUT. Events with a DUT spatial resolution significantly better than the track-position resolution are selected, and the latter is obtained from the differences between the track-positions reconstructed by the beam telescope and those by the DUT. In this section, details on the selection criteria, position-reconstruction and spatial resolution of such events are presented. Furthermore, the impact of various factors, such as electronics noise, cross-talk, and energetic $\delta$-electrons is investigated.

\begin{figure}[!ht]
\centering
\includegraphics[width=0.7\textwidth]{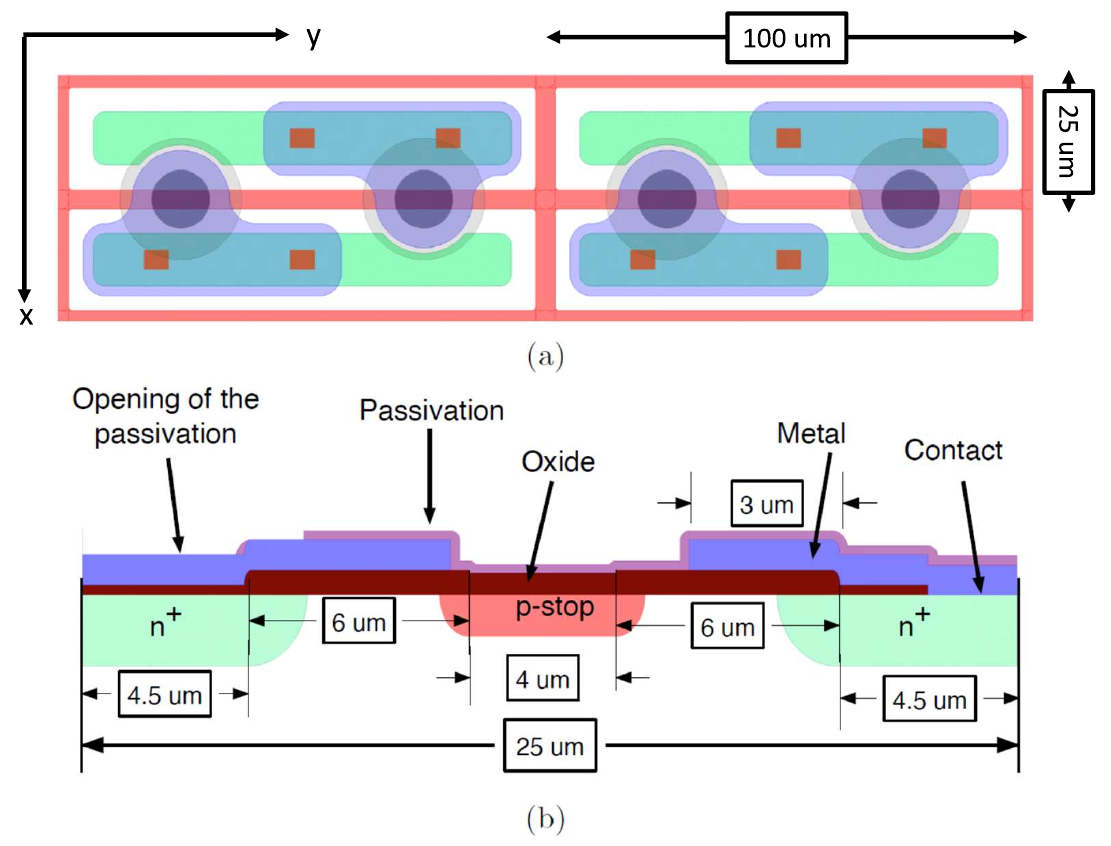}
\caption{(a)\,Top view of four pixels of the silicon pixel detector used in this paper.
(b)\,Cross section of two adjacent half pixels in the \SI{25}{\um} direction.
In the simulations the circular bond pads, which result in an asymmetric cross-talk, are not implemented. For figure colors, refer to the online version; figure taken from\,\cite{Pixel:2023}.}
\label{fig:4Pixels}
\end{figure}

The method is demonstrated using \num{1e5} events simulated with the program PIXELAV\,\cite{Swartz:2002}. The response of a \SI{150}{\um} thick silicon sensor with $\SI{25}{\um}\times \SI{100}{\um}$ pixels to \SI{40}{\GeV}/c pions, uniformly distributed over one pixel with normal incidence, was simulated. Figure\,\ref{fig:4Pixels} shows the pixel layout of the actual sensor used, which corresponds to the layout of a prototype pixel sensor for the CMS Tracker Phase-2 upgrade~\cite{Pixel:2023}. The coordinate normal to the \SI{100}{\um} width is called $x$, the one normal to the \SI{25}{\um} width, $y$, and $z$ the normal to the sensor plane. 
In the simulation the bond-pads, which have to match the RD53A bond-pattern, are not included. The bond pads of the sensor, which are positioned above every second pixel interface in the $x$-direction, are expected to cause non-negligible cross-talk due to capacitative coupling. The distance between the pixel implants is about \SI{16}{\um} in both directions. Simulation with the program of Ref.\,\cite{Ebrahimi:2021}, which assumes zero inter-pixel distance and does not include the range of $\delta $-electrons, gives compatible results.

A \emph{p}-type sensor with boron doping of \SI{4.4e12}{\per\cubic\cm}, biased at \SI{120}{\V} and operated at a temperature of \SI{20}{\celsius}, was simulated.
Electronics noise is implemented by adding a Gaussian-distributed random number with mean zero, and width $\sigma _\mathit{el} = 300$\,e, where e is the charge of the electron. Cross-talk (\emph{XT}) is simulated by multiplying the charge matrix by cross-talk matrices. Examples of how cross-talk is implemented are given later in this section. 
The total simulated charge for a given event is called $Q$. When plotted over many events, $Q$ follows the typical Landau-distribution as a probability function with a most-probable-value of about \SI{11.1}{ke}, and a mean of \SI{14.1}{ke}. To reduce the effects of $\delta$-electrons, events with $Q > \SI{16}{ke}$ are rejected. 

\begin{figure}[!ht]
\centering
\begin{subfigure}[a]{0.5\textwidth}
\includegraphics[width=\textwidth]{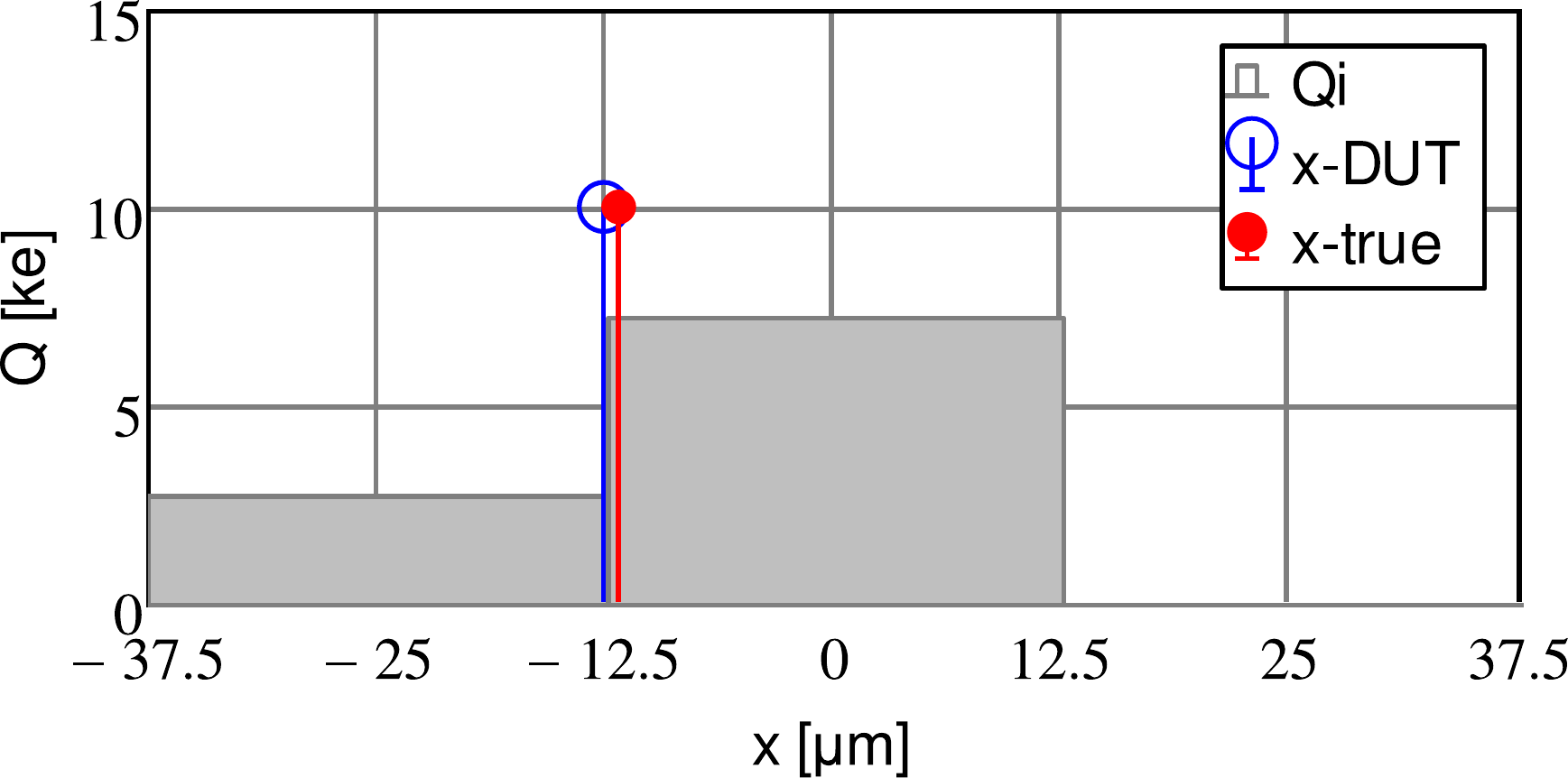}
\caption{ }
\label{fig:Evt0}
\end{subfigure}%
~
\begin{subfigure}[a]{0.5\textwidth}
\includegraphics[width=\textwidth]{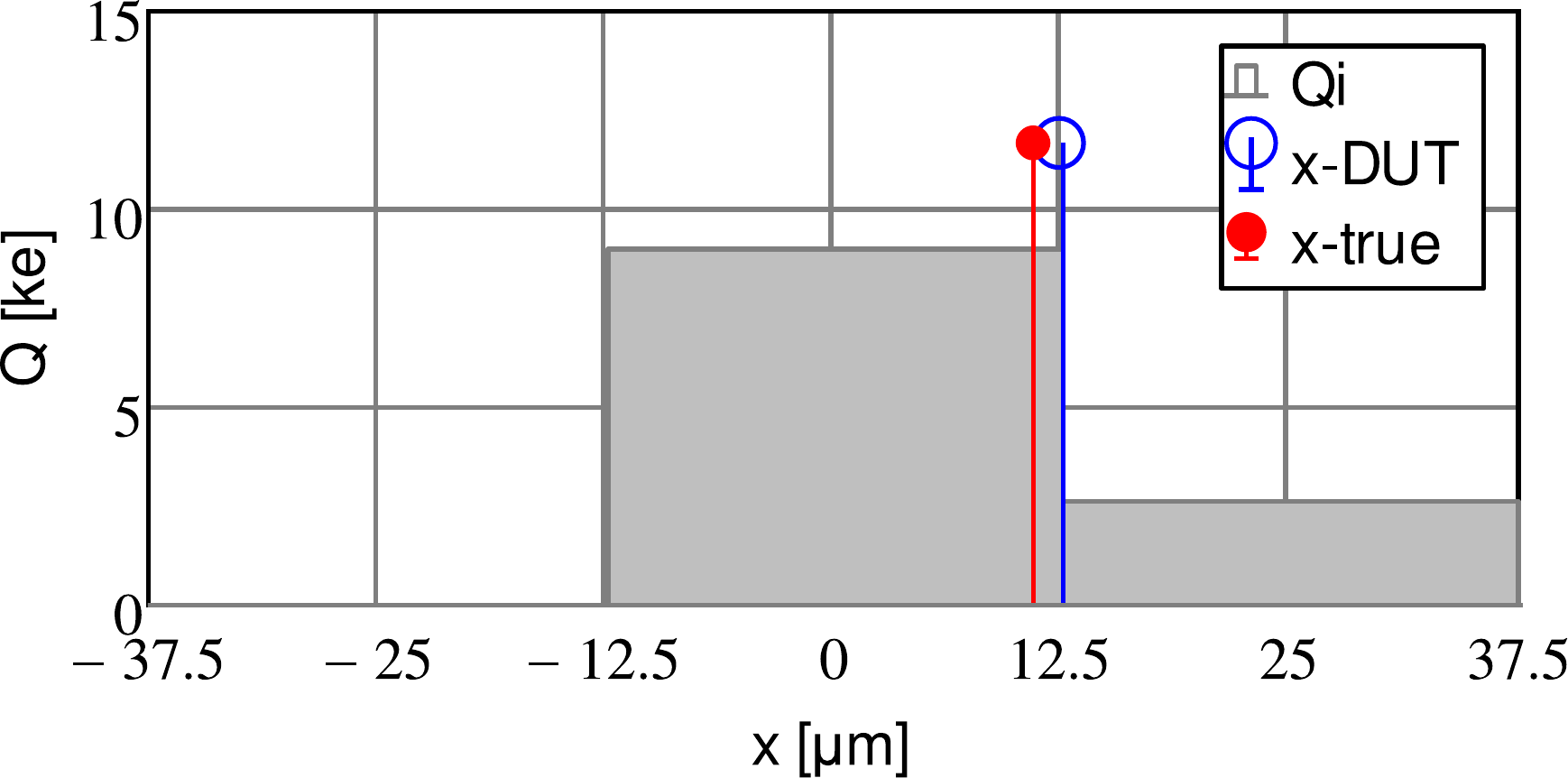}
\caption{ }
\label{fig:Evt1}
\end{subfigure}%

\caption{Two events illustrating the calculation of the position $x_\mathrm{DUT}$ for cluster-size-two~events in the \SI{25}{\um} direction. The vertical lines with the filled circle on top indicate the simulated particle positions $x_\mathit{true}$. The vertical grid lines show the pixel centres and boundaries. The shaded histograms present the simulated charges above threshold, summed over the \SI{100}{\um} direction.
(a)\,The value of $x_\mathit{true} = - \SI{11.676}{\um}$, $\eta_x = 0.483$, and the reconstructed position $x_\mathrm{DUT} = -\SI{12.50}{\um}$, corresponding to the boundary between the pixels with $Qx > 0$.
(b)\,The value of $x_\mathit{true} = +\SI{11.084}{\um}$, $\eta_x = - 0.612$, and the reconstructed position, $x_\mathrm{DUT} = \SI{12.50}{\um}$, corresponding to the boundary between the pixels with $Qx > 0$.}
\label{fig:Evt}
\end{figure}

For every event, the charge collected by each pixel is denoted as $\mathit{Qp}_\mathit{ix,jy}$, where $\mathit{i}$ and $\mathit{j}$ represent the row and column indices of the pixel in the sensor coordinate system. Therefore, pixels with the same $i$-value share the same $x$-coordinate value, and likewise for $jy$. Charge values smaller than the threshold $\mathit{Thr} = \SI{1200}{e}$, are set to zero.
A group of adjacent pixels with charge greater than zero defines a cluster.
For clusters in the $x$-direction the charges with the same \emph{ix}-values, are summed over \emph{jy} and are called \emph{Qx}$_\mathit{ix}$. The number of adjacent \emph{Qx}$_\mathit{ix}$-values greater than zero is defined as the projected \emph{x}-cluster-size, \emph{x-cls}. Next, events with \emph{x-cls}\,$=2$ are selected. The same procedure is followed for the $y$-direction. In the following, to indicate events with a projected cluster-size of two in at least one of the two directions, the term \hbox{"cluster-size-two events"} will also be used.
Figure\,\ref{fig:Evt} shows two different \emph{x-cls}\,$=2$ events.

For the events with \emph{x-cls}\,$=2$ the boundary between the two pixels with signals above the threshold is assigned to the reconstructed position and defined as $x_\mathrm{DUT}$, and the charge asymmetry
\begin{equation}\label{eq:eta}
\eta_x = \frac{Qx_2 - Qx_1} {Qx_1 + Qx_2}
\end{equation}
is calculated; $Qx_1$ is the charge in the pixels with the lower, and $Qx_2$ the one with the higher $x$-value. The \hbox{$y$-direction} is treated analogously. The asymmetry is closely related to the parameter $\eta _\mathrm{Ref} = Qx_2 / (Qx_1+Qx_2)$ introduced in Ref.\,\cite{Belau:1983}, which has been used to measure the transverse diffusion of the charge carriers drifting in the sensor, which is the physics basis of the proposed method.

\begin{figure}[!ht]
\centering
\begin{subfigure}[a]{0.5\textwidth}
\includegraphics[width=\textwidth]{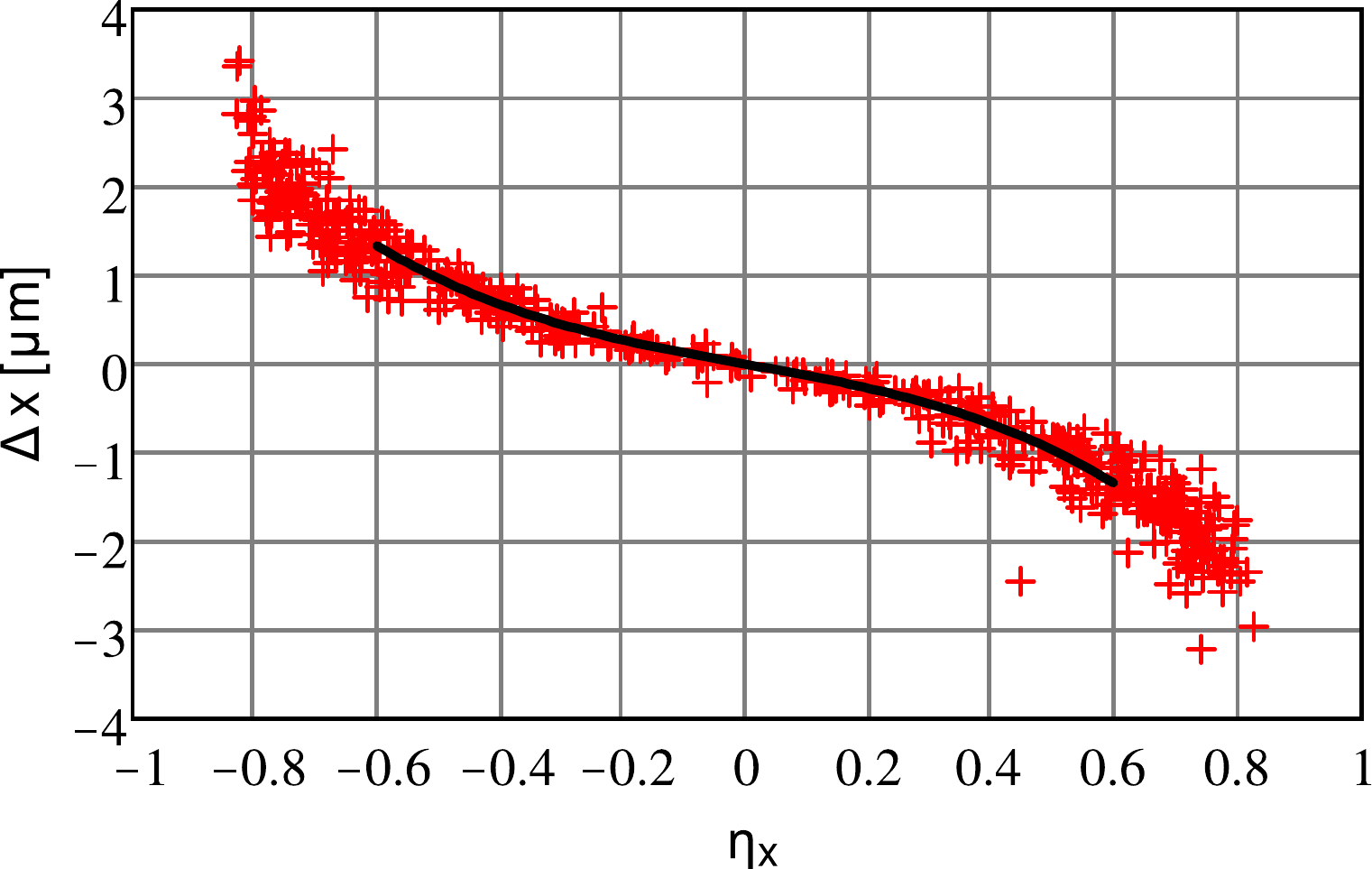}
\caption{ }
\label{fig:Dx-eta_XT0}
\end{subfigure}%
~
\begin{subfigure}[a]{0.5\textwidth}
\includegraphics[width=\textwidth]{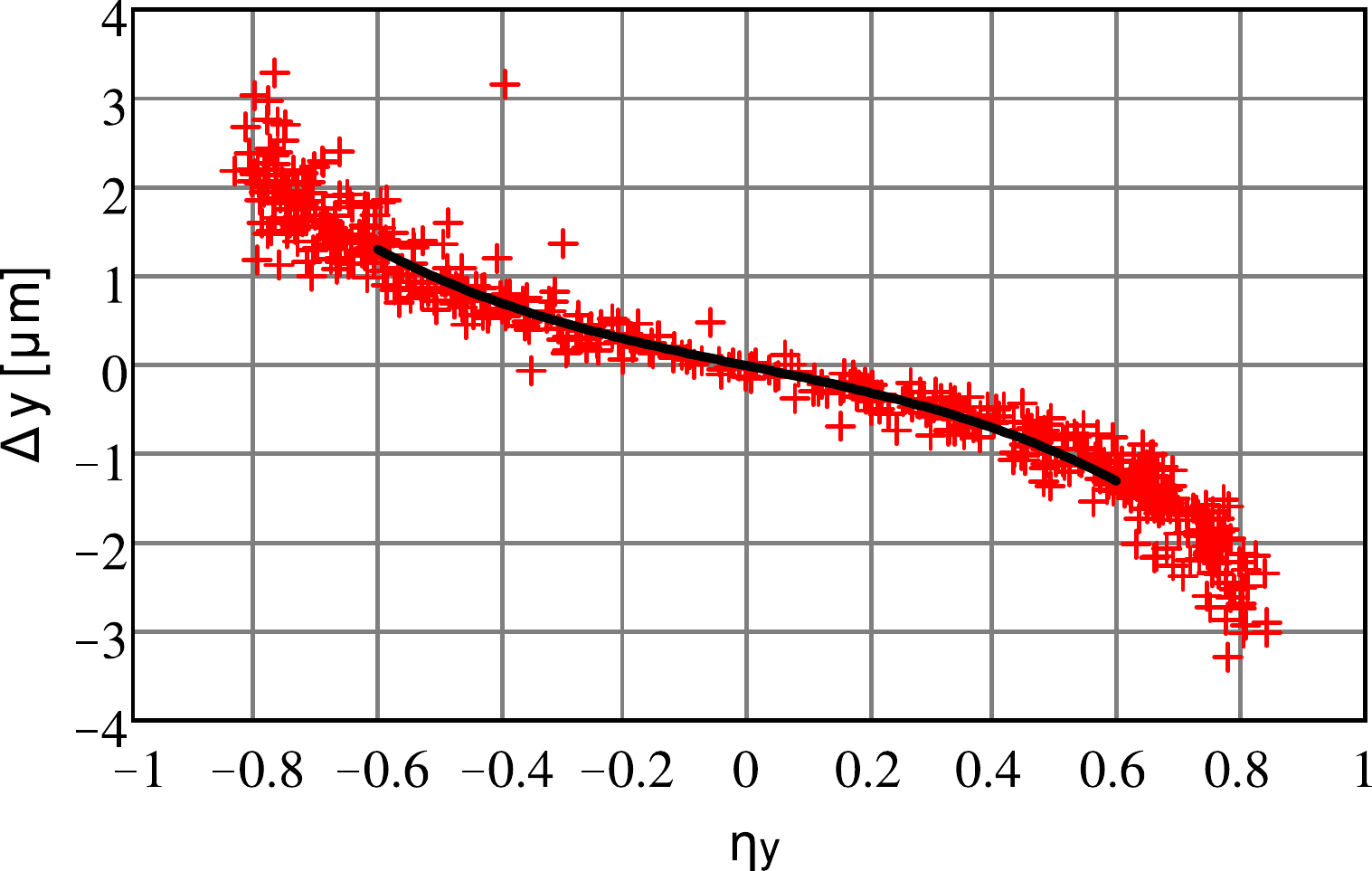}
\caption{ }
\label{fig:Dy-eta_XT0}
\end{subfigure}%
\caption{(a) Scatter plot of the residuals $\Delta x = x_\mathrm{DUT} - x_\mathit{true}$ versus $\eta_x$, and (b) $\Delta y$ versus $\eta_y$.
The fact that the boundary between the pixels is assigned to $x_\mathrm{DUT}$ and $y_\mathrm{DUT}$, whereas the true position moves towards the centre of the pixel for increasing $|\eta|$, causes the S-shape.
The solid black lines are regression of third-order polynomials for $ -0.6 < \eta < + 0.6$, which can be used to correct $x_\mathrm{DUT}$ and $y_\mathrm{DUT}$. }
\label{fig:D-eta_XT0}
\end{figure}

Figure\,\ref{fig:Dx-eta_XT0} shows a scatter plot of the residuals $\Delta x = x_\mathrm{DUT} - x_\mathit{true}$ versus $\eta_x$, and Figure\,\ref{fig:Dy-eta_XT0} of $\Delta y = y_\mathrm{DUT} - y_\mathit{true}$ versus $\eta_y$, where $x_\mathit{true}$ and $y_\mathit{true}$ are the true $x$ and $y$ track-coordinates at the DUT provided by the simulation.
Here, and in the following sections, $\Delta$ is used to represent simultaneously the residuals for both $\Delta x$ and $\Delta y$. This stylistic choice, without indices, is also used for $\eta$ and all other terms referring to $x$ and $y$ distributions at the same time.

It is noted that, in spite of the factor four difference in $x$- and $y$-pixel pitch, the scatter plots are similar and only the fractions of events differ. For the \SI{100}{\um} pitch there are approximately $100/25 = 4$ times less cluster-size two events than for the \SI{25}{\um} pitch.
For $|\eta| \approx 0$, for which the charge in the two pixels is approximately the same, the widths of the $\Delta$ distributions are less than $150\,$nm. With increasing $|\eta|$ the asymmetry increases, the true position moves inwards in the pixel with higher charge, whereas $x_\mathrm{DUT}$ and $y_\mathrm{DUT}$ remain at the pixel boundary. For $|\eta| < 0.5$ the shift is still smaller than \SI{1}{\um}. If a higher accuracy or an increase of the $|\eta|$\,range to gain statistics are required, a correction for the mean of $\Delta x$ and $\Delta y$ can be made. For example, the corrected DUT position in $x$ is $x_\mathrm{DUT,\,corr} = x_\mathrm{DUT} - \langle \Delta x (\eta_x) \rangle $, where $\langle \Delta x (\eta_x) \rangle $ is the mean of the $\Delta x (\eta_x)$ distribution shown in Figure\,\ref{fig:Dx-eta_XT0}. Regressions by third-order polynomials in the range $-0.6 < \eta_x < +0.6$ were used to obtain a parametrization for $\langle \Delta x (\eta_x ) \rangle $. They are shown as black solid lines in the figure. In an alternative method for determining the correction, $\eta_x$ is divided into intervals and the median of $\Delta x (\eta_x)$ is calculated. In this way the sensitivity to outliers is reduced. An example is shown in Figure\,\ref{fig:D-eta_non}. For the $y$-direction the same procedure is applied.

\begin{figure}[!ht]
\centering
\begin{subfigure}[a]{0.5\textwidth}
\includegraphics[width=\textwidth]{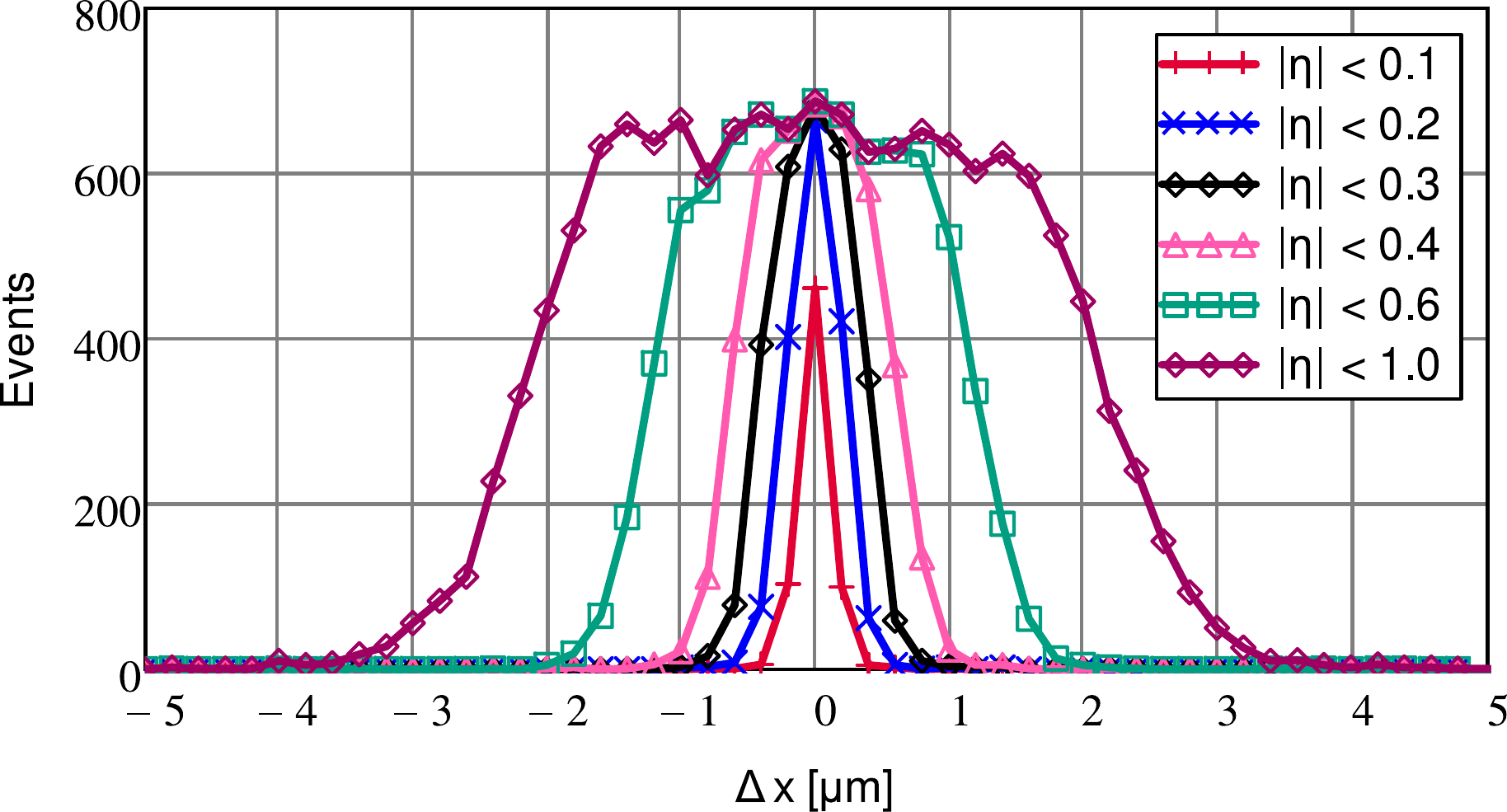}
\caption{ }
\label{fig:Dx_XT0}
\end{subfigure}%
~
\begin{subfigure}[a]{0.5\textwidth}
\includegraphics[width=\textwidth]{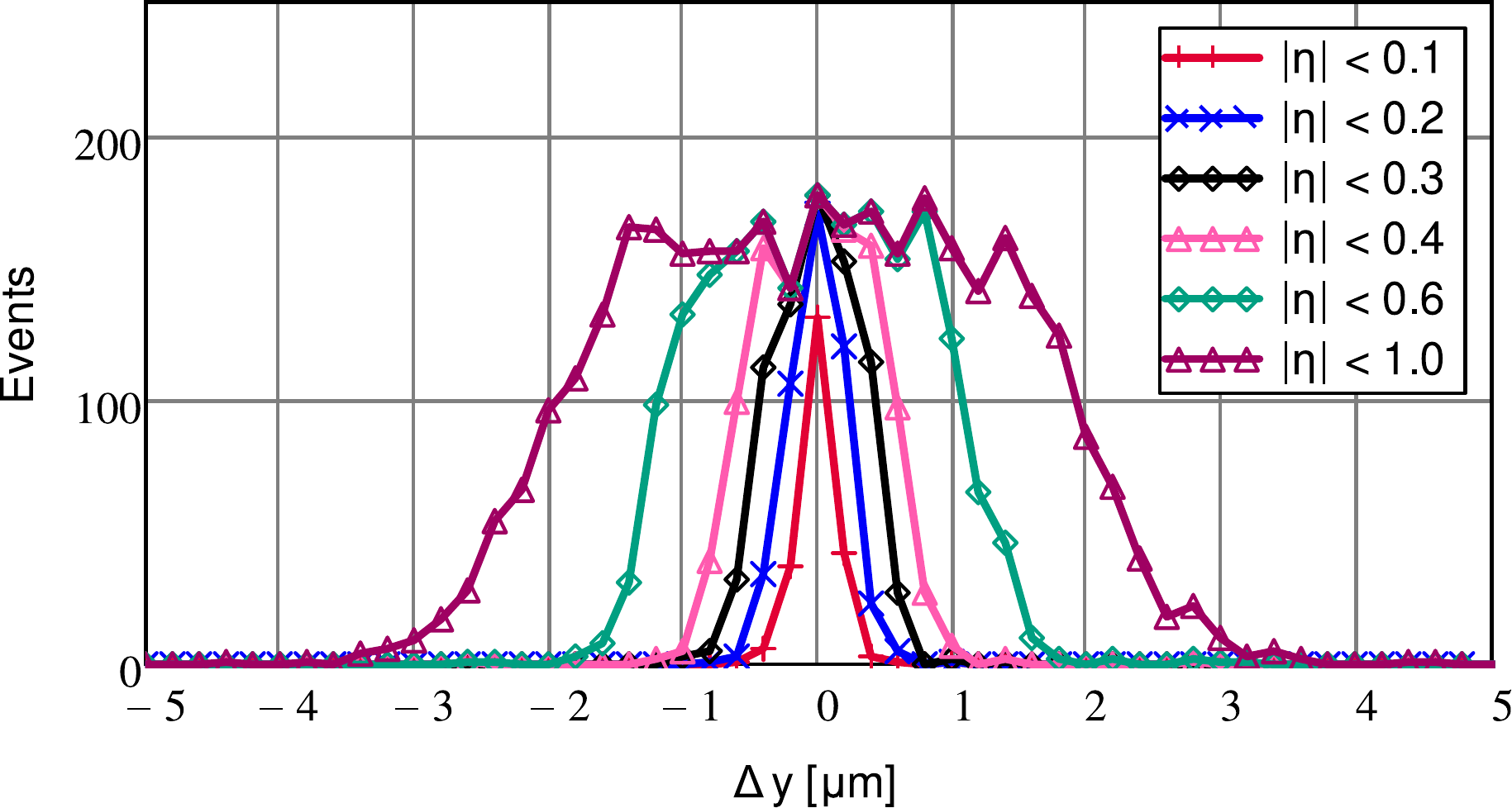}
\caption{ }
\label{fig:Dy_XT0}
\end{subfigure}%
\caption{Residuals distributions $\Delta x = x_\mathrm{DUT} - x_\mathit{true}$ and $\Delta y = y_\mathrm{DUT} - y_\mathit{true}$ of the cluster-size-two~events for different $|\eta|$\,cuts,
(a)\,for the \SI{25}{\um} direction (\emph{x}), and
(b)\,for the \SI{100}{\um} direction (\emph{y}).}
\label{fig:D_XT0}
\end{figure}

The correction is related to the diffusion of the charge carriers during their drift in the sensor and, as discussed in Ref.~\cite{Belau:1983}, it depends on the sensor thickness, the doping concentration and the applied voltage. For irradiated sensors, where charges are trapped, the knowledge of the trapping distance of the charge carriers and of the weighting field is required in addition. Figure\,\ref{fig:D_XT0} shows the distribution of the residuals $\Delta x$ and $\Delta y$ for the cluster-size-two~events and different $|\eta|$\,cuts. Apart from the factor 4 reduction of events, the distributions for the $x$- and $y$-directions are similar. When the value of the $|\eta|$\,cut is increased, events are added at higher $|\Delta|$\,values. As expected, the products of pixel pitch times the fraction of events after the $|\eta|$\,cuts agree with the full-width-half-maximum of the $\Delta$\,distributions.

\begin{figure}[!ht]
\centering
\begin{subfigure}[a]{0.5\textwidth}
\includegraphics[width=\textwidth]{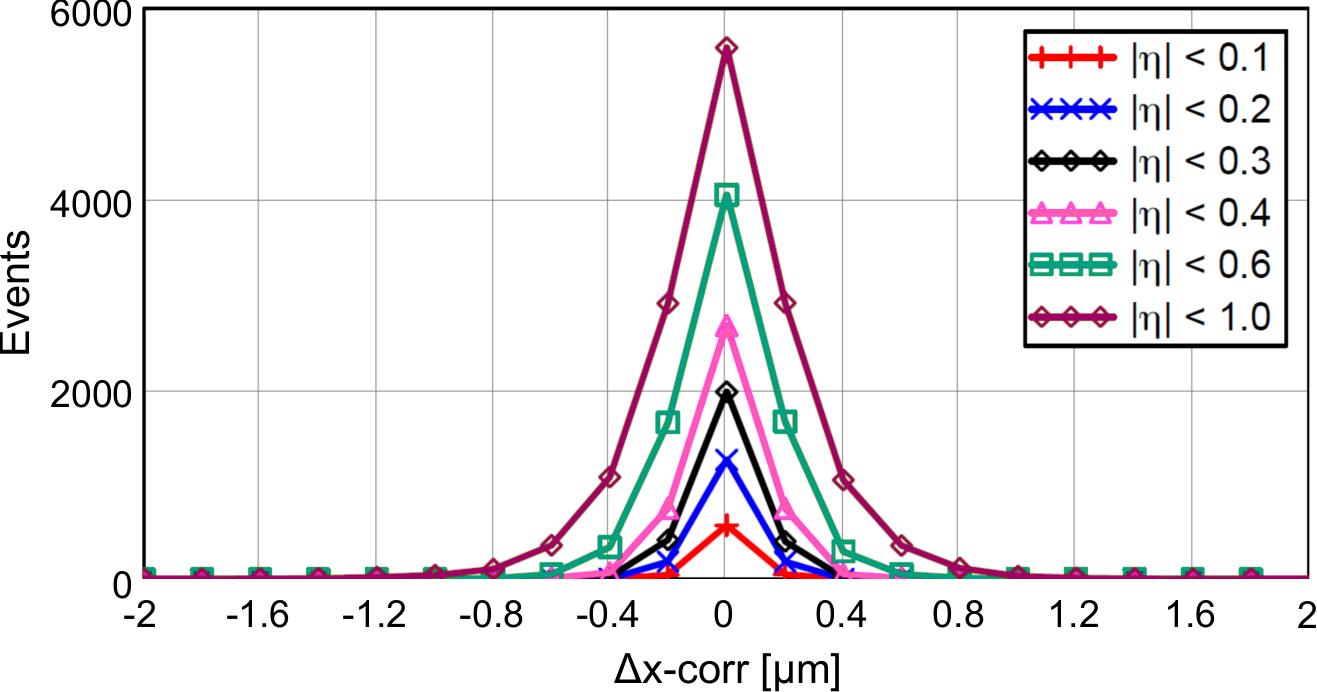}
\caption{ }
\label{fig:Dxcorr_XT0}
\end{subfigure}%
~
\begin{subfigure}[a]{0.5\textwidth}
\includegraphics[width=\textwidth]{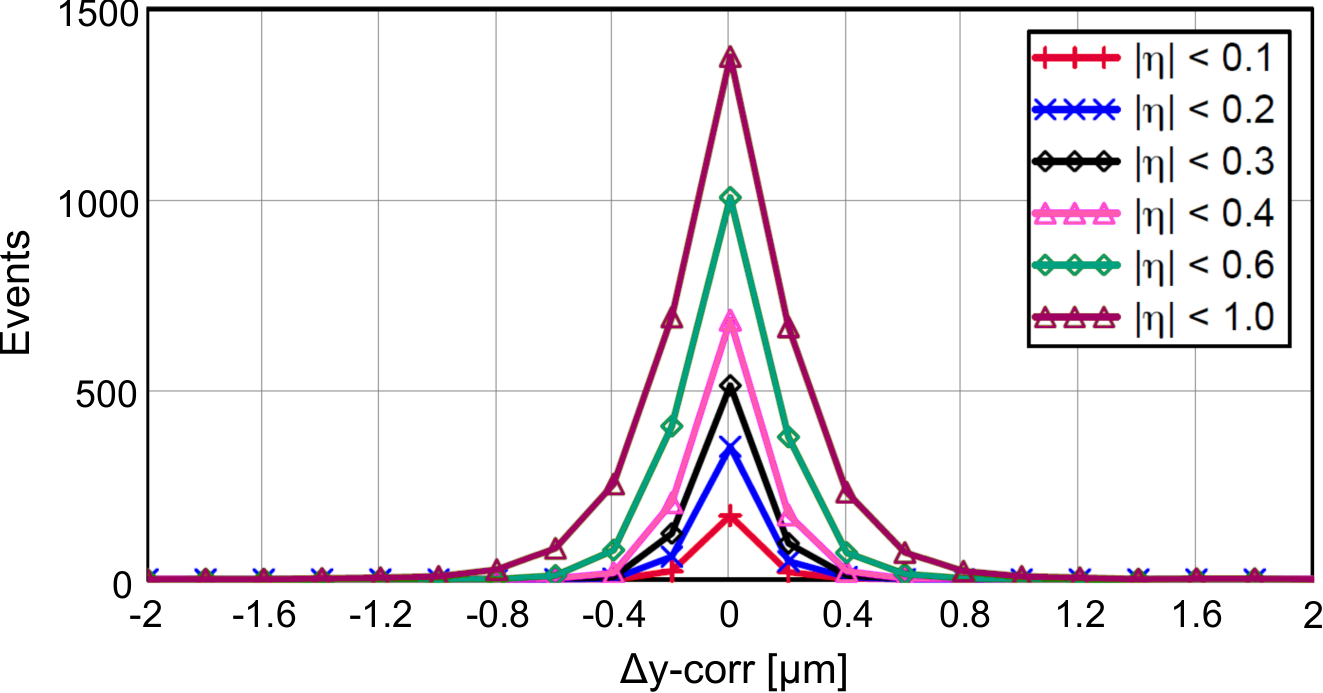}
\caption{ }
\label{fig:Dycorr_XT0}
\end{subfigure}%
\caption{Residuals distributions $\Delta x\text{\small -corr} = x_\mathrm{DUT,corr} - x_\mathit{true}$ and $\Delta y\text{\small -corr} = y_\mathrm{DUT,corr} - y_\mathit{true}$ of the cluster-size-two~events for different $|\eta|$\,cuts,
(a)\,for the \SI{25}{\um} direction (\emph{x}), and
(b)\,for the \SI{100}{\um} direction (\emph{y}).
The $x_\mathrm{DUT,corr}$ and $y_\mathrm{DUT,corr}$ positions are obtained by subtracting for a given $\eta$ the values of the third-order regression shown in Figure\,\ref{fig:D-eta_XT0} from the DUT\,positions.}
\label{fig:Dcorr_XT0}
\end{figure}

Figure\,\ref{fig:Dcorr_XT0} shows the distributions of the residuals when the correction is applied, defined as $\Delta x\text{\small -corr} = x_\mathrm{DUT,corr} - x_\mathit{true}$ and $\Delta y\text{\small -corr} = y_\mathrm{DUT,corr} - y_\mathit{true}$, respectively.

\begin{figure}[!ht]
\centering
\begin{subfigure}[a]{0.5\textwidth}
\includegraphics[width=\textwidth]{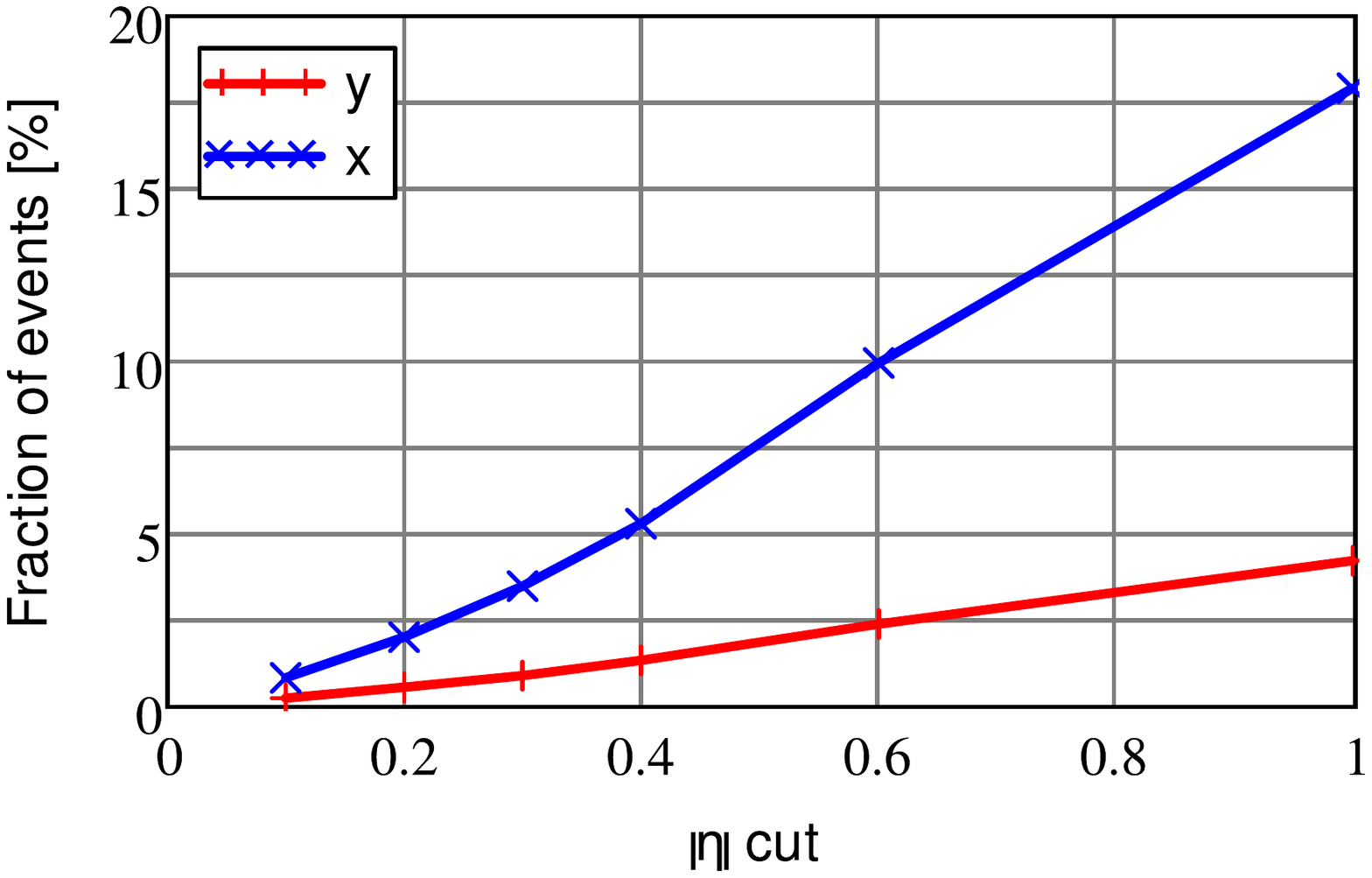}
\caption{ }
\label{fig:fr-etaXT0}
\end{subfigure}%
~
\begin{subfigure}[a]{0.5\textwidth}
\includegraphics[width=\textwidth]{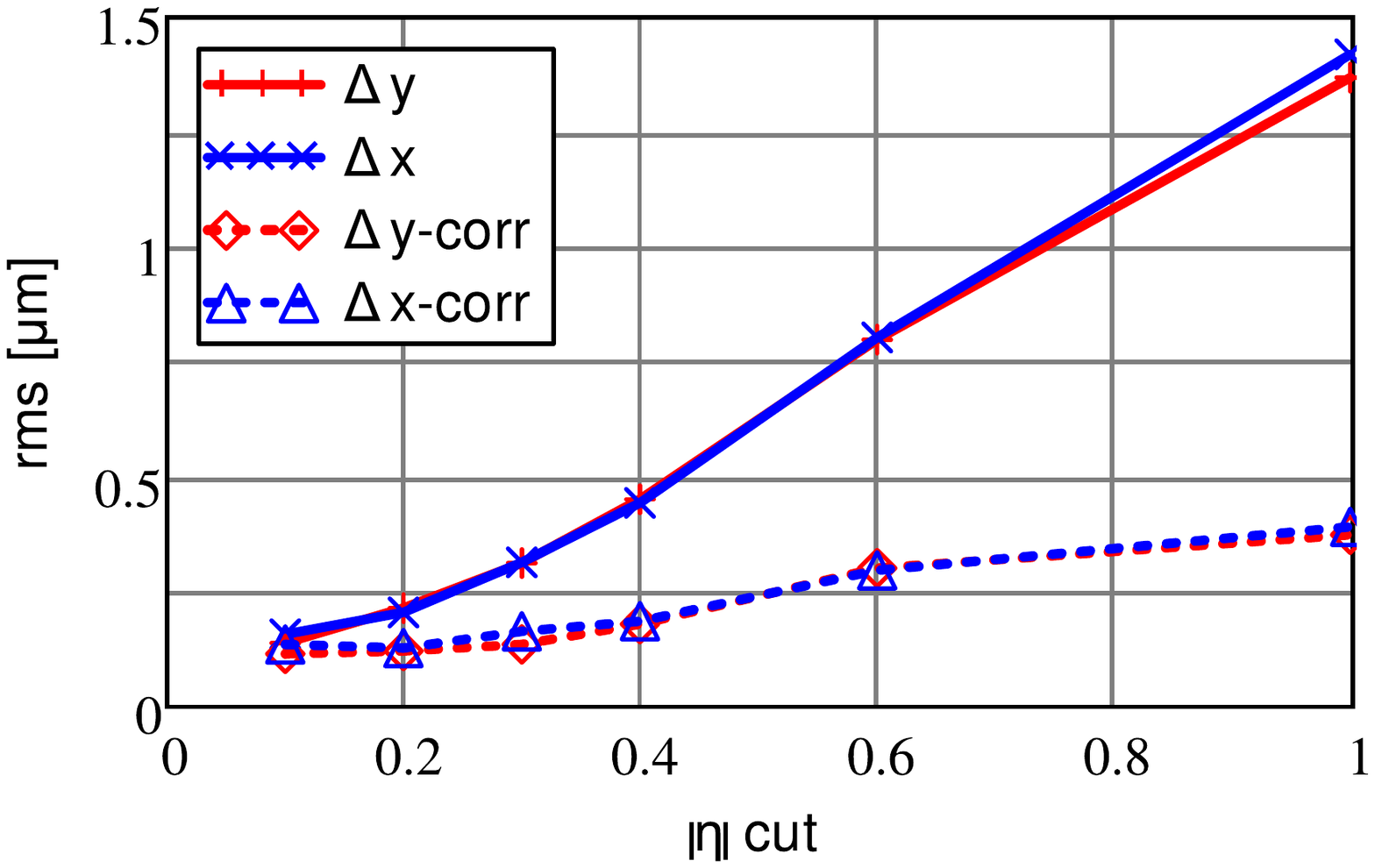}
\caption{ }
\label{fig:Rms-etaXT0}
\end{subfigure}%
\caption{(a) Fraction of events with cluster-size-two as a function of the $|\eta|$\,cut.
(b) \emph{rms} of the residual distributions $\Delta$ and $\Delta \text{\small -corr}$ for cluster-size-two~events as a function of the $|\eta|$\,cut.}
\label{fig:frRms-XT0}
\end{figure}

The fractions of events and the \emph{rms} (root-mean-square) values as a function of the $|\eta |$\,cut are shown in Figure\,\ref{fig:frRms-XT0}. The fraction of events increases approximately linearly with the $|\eta |$\,cut, and the ratio of events in the $x$ to the one in the $y$-direction agrees with the inverse of the pixel pitches. The \emph{rms} values for $x$ and $y$ agree. Whereas for the DUT the \emph{rms} increases linearly with the $|\eta |$\,cut, reaching a value of about \SI{0.5}{\um} for $|\eta| < 0.5$, the increase for DUT$_\mathit{corr}$ is significantly less: below \SI{0.4}{\um} for the entire $|\eta |$ range. 
It is concluded that the proposed method should be able to precisely determine the resolution of the tracks reconstructed in a beam telescope and extrapolated to the DUT. In addition, the dependence of the measured resolution on the $|\eta |$\,cut provides an important cross-check.

\subsection{Cross-talk}
\label{subsect:Cross-talk}
\begin{figure}[!ht]
\centering
\begin{subfigure}[a]{0.5\textwidth}
\includegraphics[width=\textwidth]{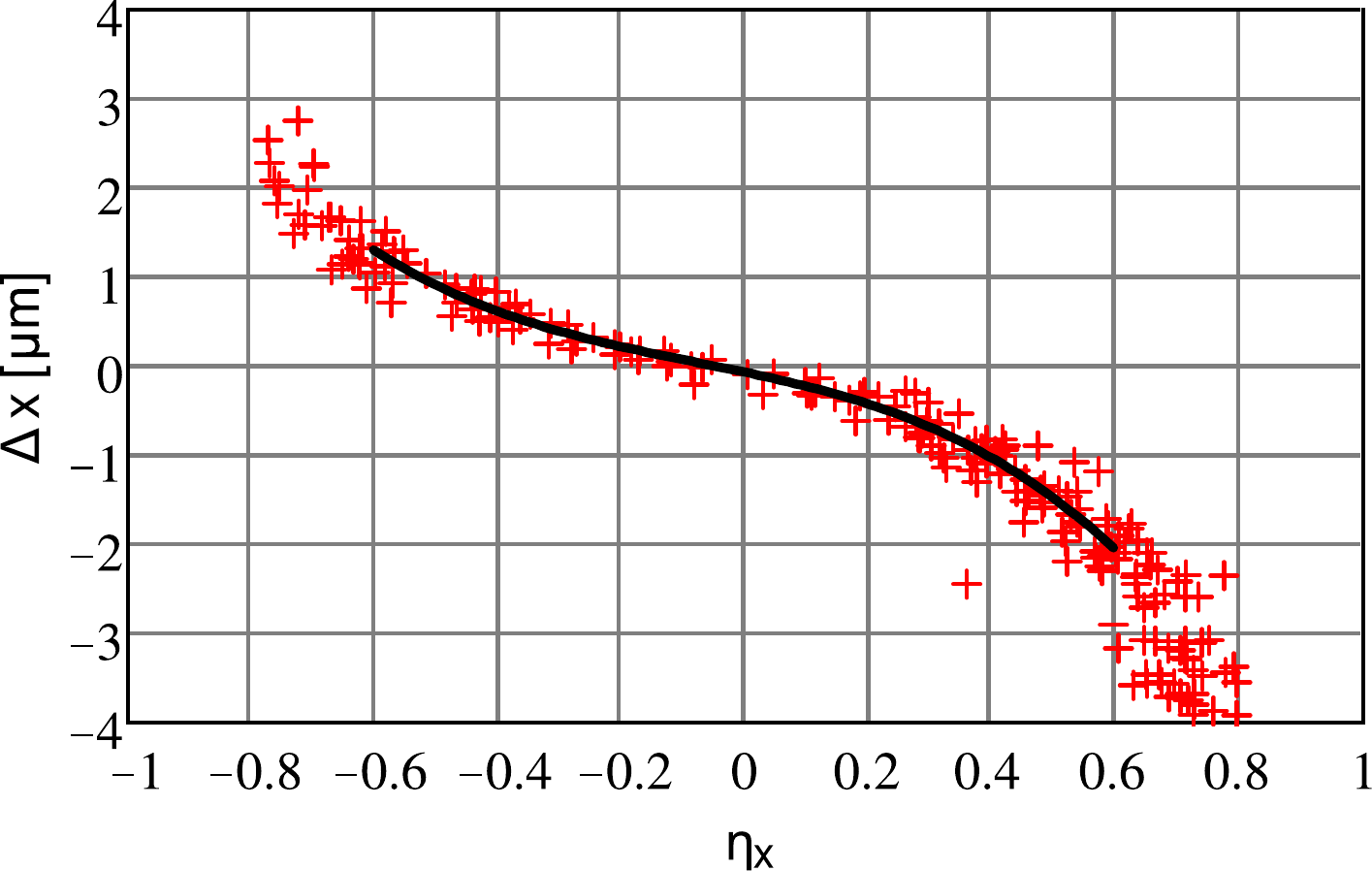}
\caption{ }
\label{fig:Dx-eta_XT10.pdf}
\end{subfigure}%
~
\begin{subfigure}[a]{0.5\textwidth}
\includegraphics[width=\textwidth]{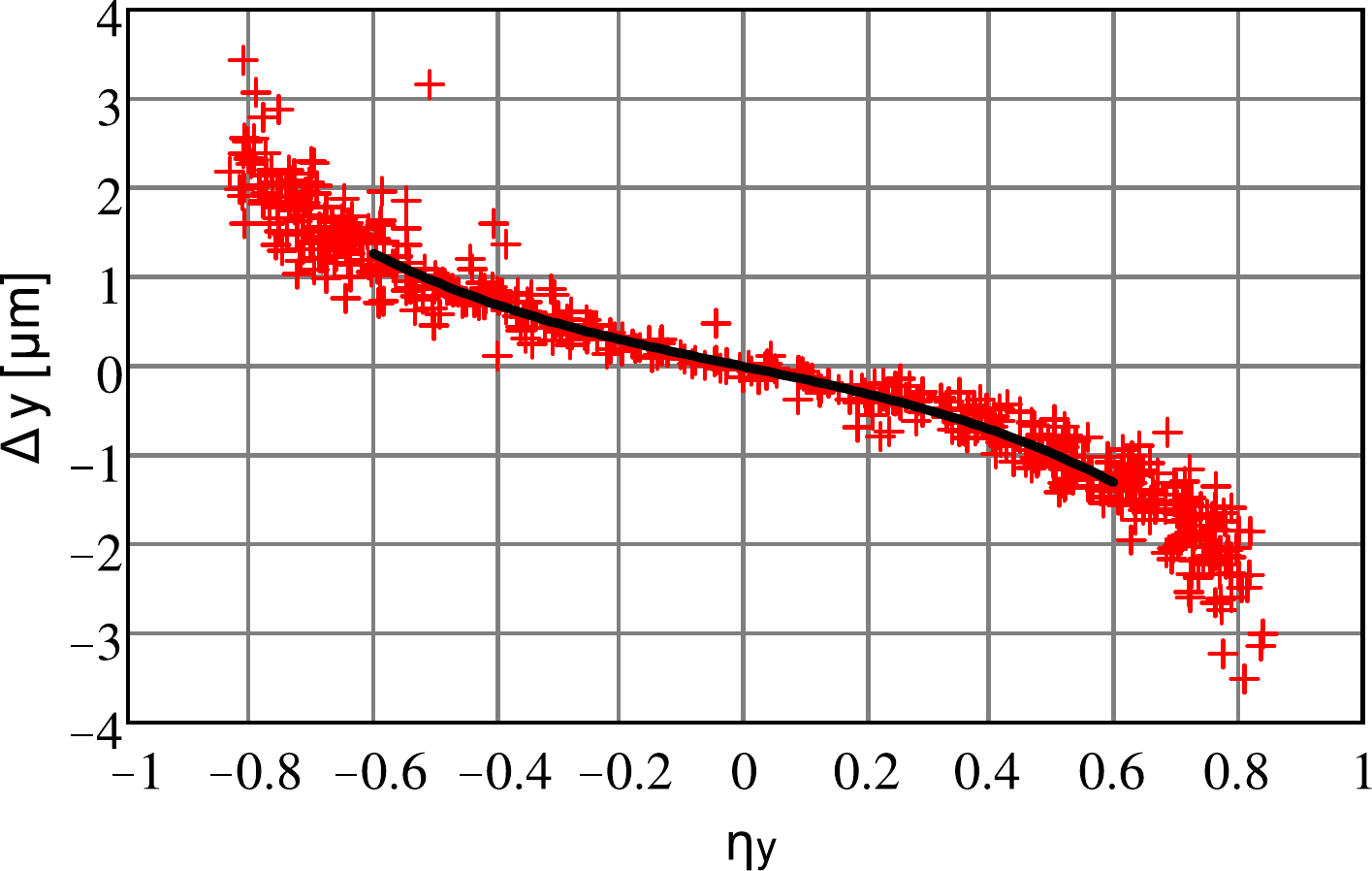}
\caption{ }
\label{fig:Dy-eta_XT10}
\end{subfigure}%
\caption{(a) Scatter plot of the residuals $\Delta x = x_\mathrm{DUT} - x_\mathit{true}$ as a function of $\eta_x$, and (b) $\Delta y$ as a function of $\eta_y$, both with 10\,\% cross-talk in the \SI{25}{\um}\,direction (\emph{x}).
The continuous curves are regressions by  third-order polynomials for $-\,0.6 < \eta < +\,0.6$.}
\label{fig:D-eta_XT10}
\end{figure}

In the following, the influence of cross-talk (\emph{XT}) between pixels on the determination of the cluster-size-two resolution is investigated. As can be seen from the sensor design shown in Figure\,\ref{fig:4Pixels}, a non-negligible cross-talk is expected in the $x$-direction and significantly less in the $y$-direction. Cross-talk in the $x$-direction is implemented by multiplying for every \emph{jy} the cross-talk matrix \emph{Ax} with the charge\,values of the individual pixels, $\mathit{Qp_{ix,\,jy}}$, to obtain $\mathit{Qp'_{ix,\,jy}}$. Results on the cross-talk towards pixels with lower \emph{ix}\,values will be presented. In this case  the diagonal elements of \emph{Ax} are 1 -- \emph{XT}, the elements to the right of the diagonal are \emph{XT} and all other elements are zero. Next, the $\mathit{Qp'_{ix,\,jy}}$\,values exceeding the threshold, \emph{Thr}, are summed over \emph{jy} to obtain $\mathit{Qx'_{ix}}$. The remainder of the analysis is the same as for the situation without cross-talk. For the results shown below, $\mathit{XT} = 0.1$ and $\mathit{Thr} = 1200$\,e are chosen. For the cross-talk matrix in the $y$-direction, \emph{Ay}, a unit matrix is assumed, which corresponds to no cross-talk.
Figure\,\ref{fig:D-eta_XT10} shows scatter plots of $\Delta x = x_\mathrm{DUT} - x_\mathit{true}$ versus $\eta_x$ and of $\Delta y = y_\mathrm{DUT} - y_\mathit{true}$ versus $\eta_y$. 

\begin{figure}[!ht]
\centering
\begin{subfigure}[a]{0.5\textwidth}
\includegraphics[width=\textwidth]{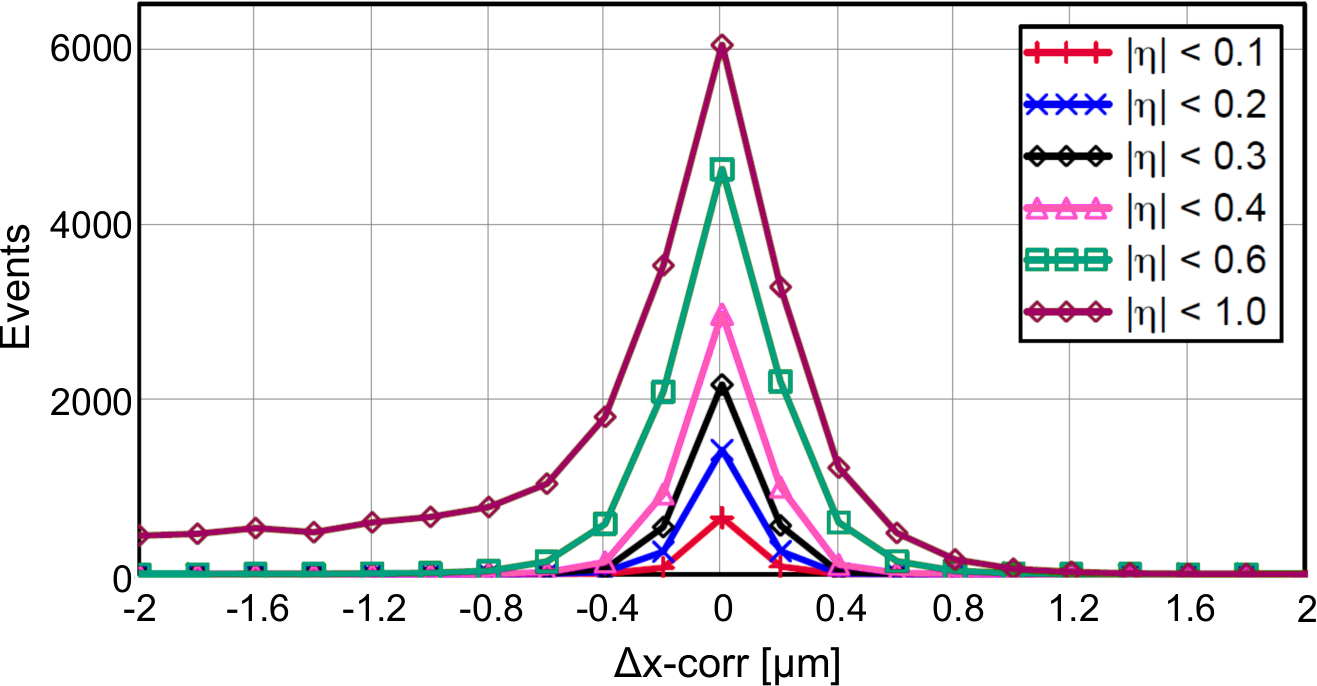}
\caption{ }
\label{fig:Dxcorr_XT10-fine}
\end{subfigure}%
~
\begin{subfigure}[a]{0.5\textwidth}
\includegraphics[width=\textwidth]{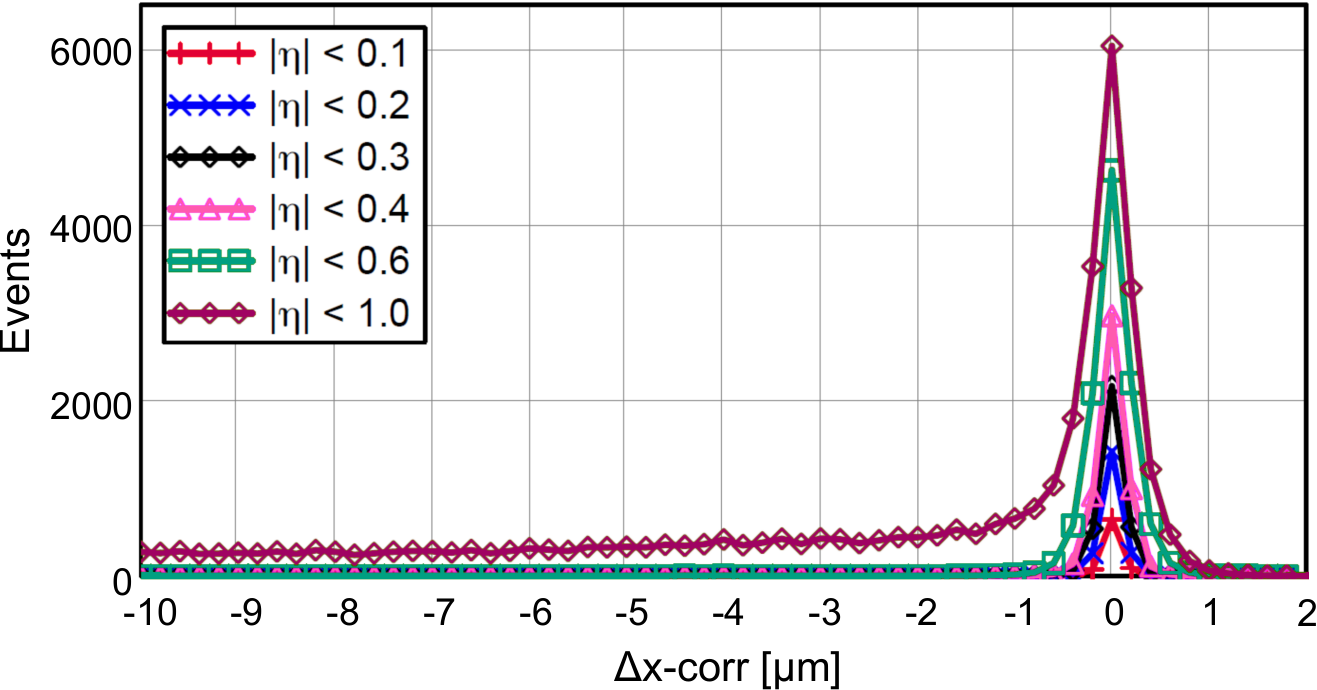}
\caption{ }
\label{fig:Dxcorr_XT10-coarse}
\end{subfigure}%
\caption{Residuals distributions $\Delta x\text{\small -corr} = x_\mathrm{DUT,corr} - x_\mathit{true}$ for different $|\eta_x|$\,cuts and 10\,\% cross-talk in the \SI{25}{\um}\,direction (\emph{x}).
(a) and (b) show the same distributions, only the $\Delta x\text{\small -corr}$ ranges differ.}
\label{fig:Dxcorr_XT10}
\end{figure}

The $\Delta y$\,distribution is similar to the one shown in Figure\,\ref{fig:Dy-eta_XT0}, and thus not affected by the 10\,\% cross-talk in the $x$-direction. However, the $\Delta x$\,distribution is affected by the cross-talk. The fraction of events with $\Delta x < - \SI{3}{\um}$ is significantly higher and $\langle \Delta x (\eta) \rangle$ is shifted to lower values. This shift is essentially negligible for negative $\eta _x$\,values, however, becomes significant for high positive $\eta _x$\,values. The first effect is caused by genuine cluster-size-one~events, which become fake cluster-size-two~events by cross-talk. The cluster-size-one~events are approximately uniformly distributed in $x$, and, for cross-talk towards pixels with lower indices, cause a flat negative $\Delta x$\,distribution extending to $\approx - \SI{25}{\um}$, the $x$-pixel pitch. This effect is illustrated by the $\Delta x\text{\small -corr}$\,distribution for $|\eta _x | < 1.0$ shown in Figure\,\ref{fig:Dxcorr_XT10}. The second effect is caused by cluster-size-two~events which remain cluster-size-two~events but with a changed $\eta _x$\,value. The magnitude of the $\eta _x$ shift can be understood with the help of Figure\,\ref{fig:Evt}. For $\eta _x < 0$, $Qx_2 < Qx_1$ (Figure\,\ref{fig:Evt1}), and the shift caused by cross-talk to lower pixel indices is small. For $\eta _x > 0$, $Qx_2 > Qx_1$ (Figure\,\ref{fig:Evt0}), and the shift, in particular for high $\eta _x$\,values, is bigger.
Figure\,\ref{fig:Dxcorr_XT10} shows how the $|\eta _x|$\,cut influences the $\Delta x\text{\small -corr}$\,distributions for 10\,\% cross-talk. Without $\eta_x$\,cut $ (|\eta _x| < 1 $) there is a long tail of negative $\Delta x\text{\small -corr}$\,values. Its origin has been discussed above. For $|\eta _x| < 0.6 $, the tail is absent and the narrow distributions without cross-talk are recovered. 

\begin{figure}[!ht]
\centering
\begin{subfigure}[a]{0.5\textwidth}
\includegraphics[width=\textwidth]{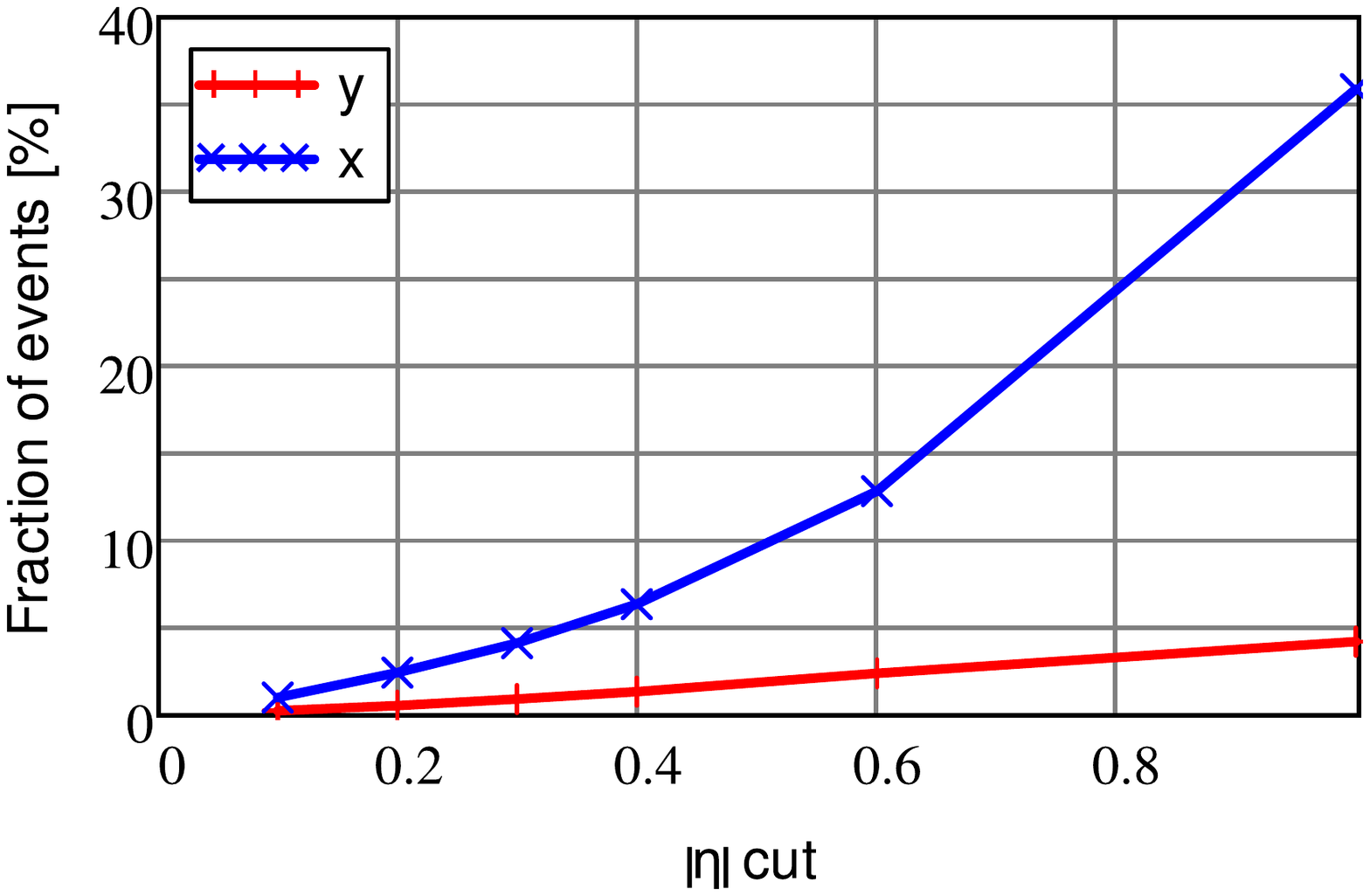}
\caption{ }
\label{fig:fr-etaXT10}
\end{subfigure}%
~
\begin{subfigure}[a]{0.5\textwidth}
\includegraphics[width=\textwidth]{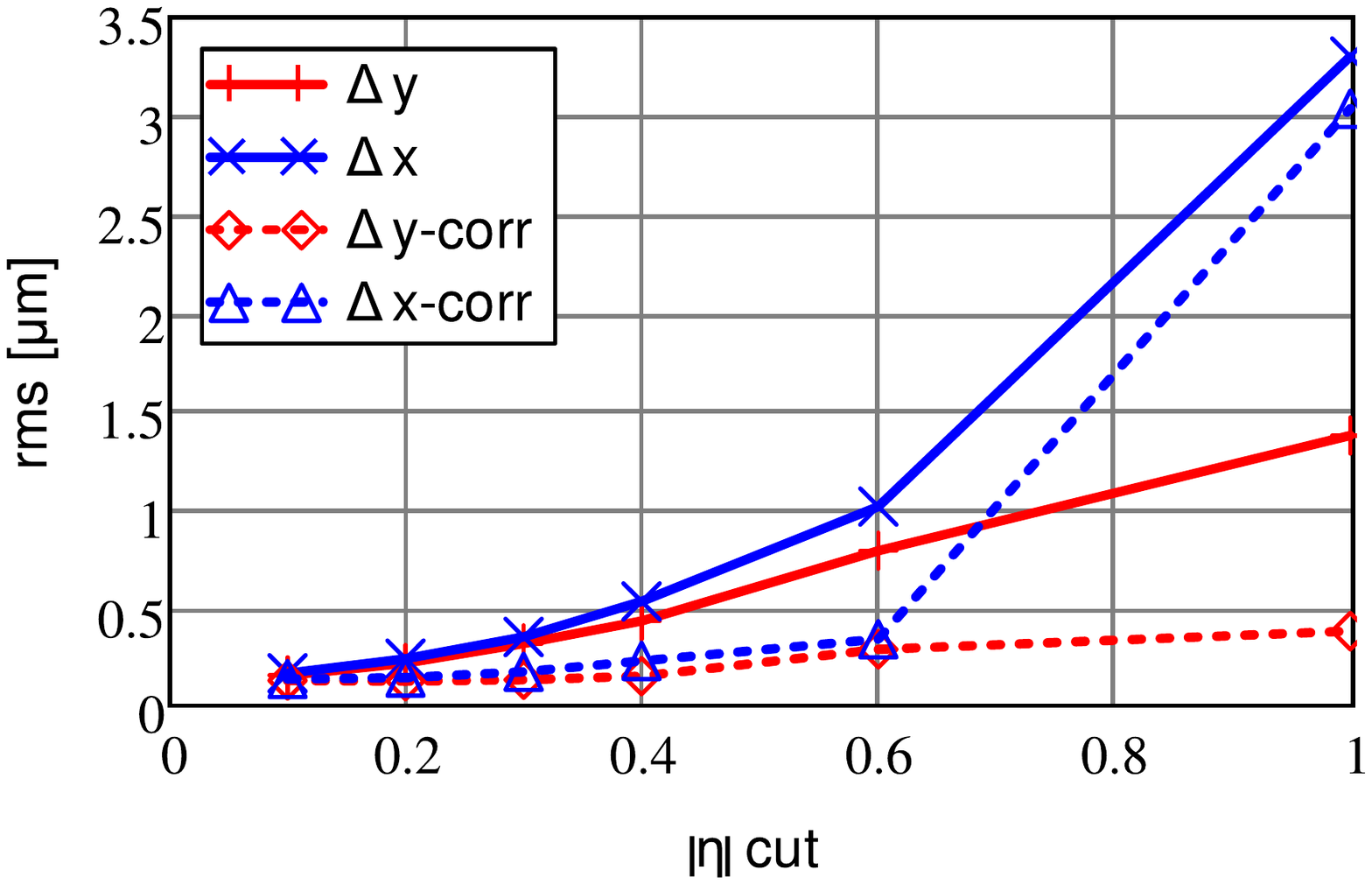}
\caption{ }
\label{fig:Rms-etaXT10}
\end{subfigure}%
\caption{(a) Fraction of events with cluster-size-two as function of the $|\eta|$\,cut for 10\,\% cross-talk in the \SI{25}{\um}\,direction (\emph{x}).
(b) \emph{rms} of the residual distributions $\Delta$ and $\Delta \text{\small -corr}$ for cluster-size-two~events as a function of the $|\eta|$\,cut for 10\,\% cross-talk in the \SI{25}{\um}\,direction (\emph{x}).}
\label{fig:frRms-XT10}
\end{figure}

Finally, in Figure\,\ref{fig:frRms-XT10} the dependence of the fraction of events and of the \emph{rms} of the $\Delta x$, $\Delta x\text{\small -corr}$, $\Delta y$\ and $\Delta y\text{\small -corr}$\,distributions on the $|\eta _x|$ and $|\eta _y|$\,cut are shown. It can be seen that without $\eta _x$\,cut, about 40\,\% of the events have an \emph{x-cls}\,$ = 2$, which is about twice the value without cross-talk in the $x$-direction. Thus, half of the events are genuine cluster-size-one~events and the remaining ones are caused by cross-talk. As expected, the cross-talk in the $x$-direction does not influence the number of cluster-size-two~events in the $y$-direction. Figure\,\ref{fig:Rms-etaXT10} shows the \emph{rms}\,values of the residual distributions as a function of the $|\eta $| cuts, which can be compared to the corresponding values without cross-talk in the $x$-direction of Figure\,\ref{fig:Rms-etaXT0}. The values for $|\eta| < 0.6$ agree. Only without $\eta_x$\,cut ($|\eta_x| < 1$) the fake cluster-size-two~events spoil the resolution in \emph{x}. It is concluded that also sensors with significant cross-talk can be used for the measurement of the track-position resolution.

\subsection{Electronics noise}
\label{subsect:Electronics noise}
\begin{figure}[!ht]
\centering
\begin{subfigure}[a]{0.5\textwidth}
\includegraphics[width=\textwidth]{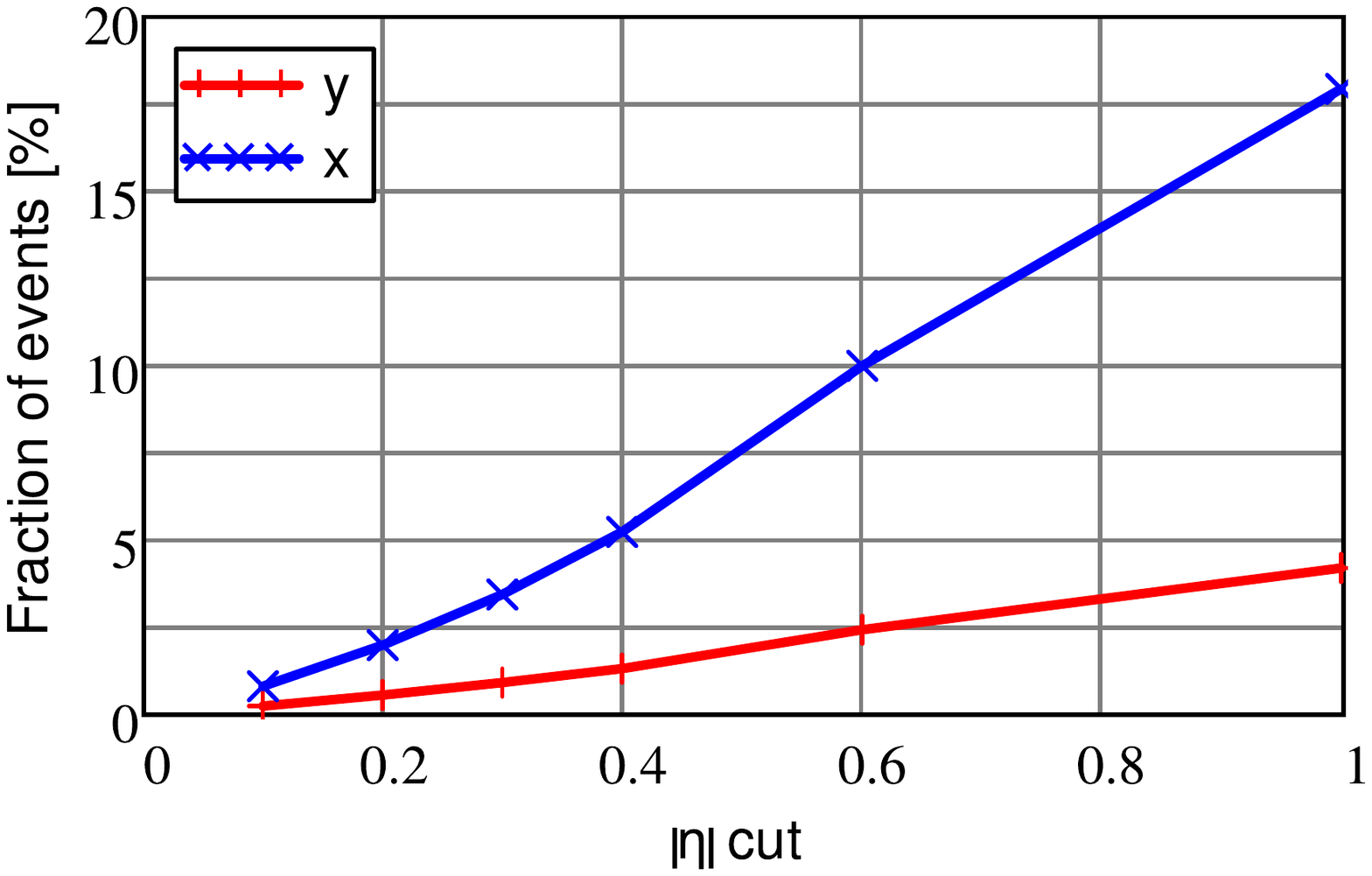}
\caption{ }
\label{fig:fr-etaNoise600}
\end{subfigure}%
~
\begin{subfigure}[a]{0.5\textwidth}
\includegraphics[width=\textwidth]{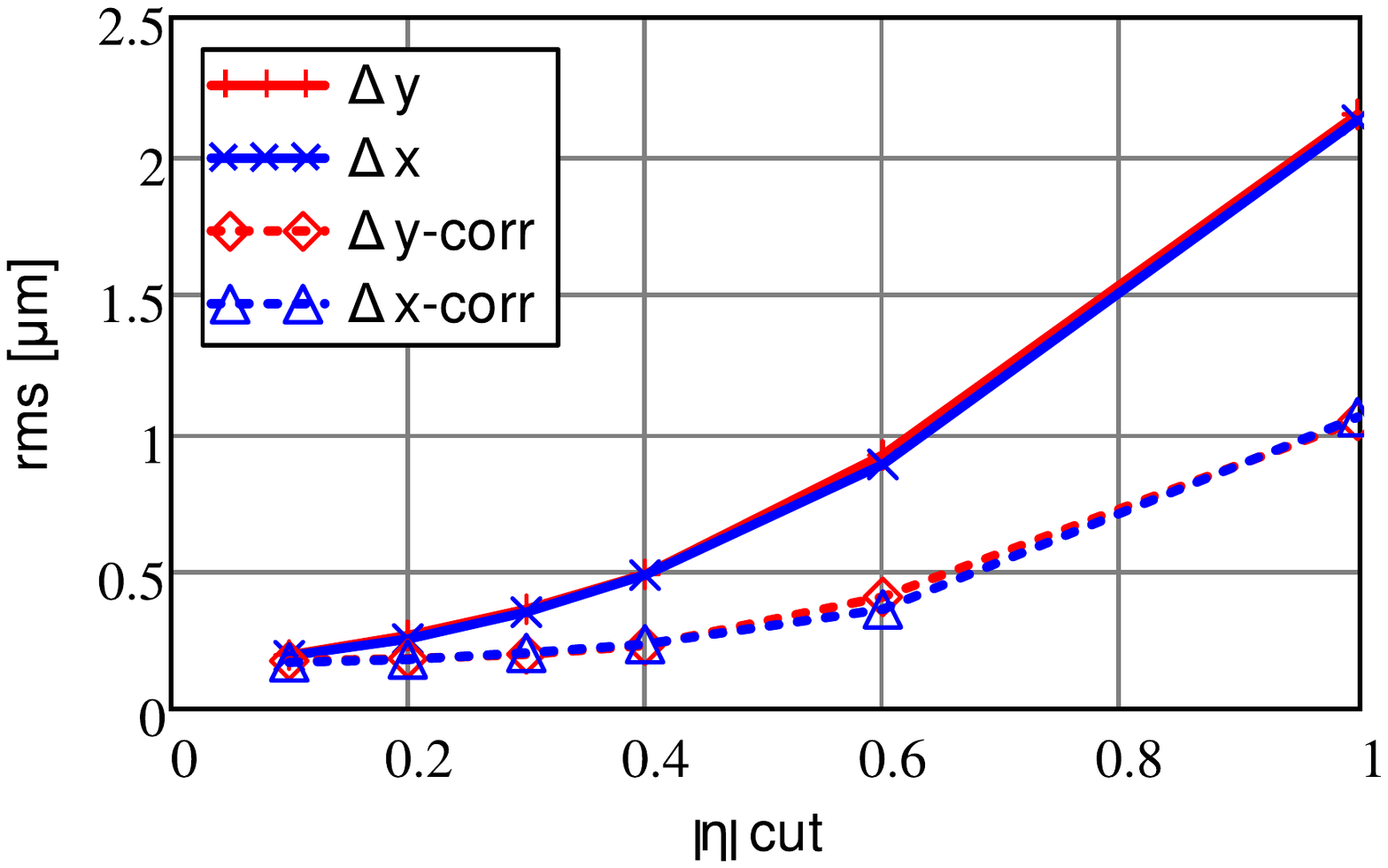}
\caption{ }
\label{fig:Rms-etaNoise600}
\end{subfigure}%
\caption{Corresponds to Figure\,\ref{fig:frRms-XT0} but with the electronics noise increased from 300\,e to 600\,e.
(a) Fraction of events with cluster-size-two as function of the $|\eta|$\,cut, and
(b) \emph{rms} of the residual distributions $\Delta$ and $\Delta \text{\small -corr}$ for cluster-size-two~events as a function of the $|\eta|$\,cut.}
\label{fig:frRms-Noise600}
\end{figure}

For the study of the influence of electronics noise on the position accuracy of cluster-size-two events, the $\sigma _\mathit{el}$ of the Gaussian is increased in the simulation from 300\,e to 600\,e. Given that the threshold for pixel hits is 1200\,e, a larger noise would result in an unacceptably high noise occupancy. Figure\,\ref{fig:frRms-Noise600} summarises the results. 

Compared to the situation with $\sigma _\mathit{el} = 300$\,e, the fraction of cluster-size-two~events increases slightly, and the \emph{rms} values of the residual distributions increase by a factor $\approx$\,1.7. However, they typically are less than \SI{1}{\um} and it can be concluded that electronics noise has no relevant influence on the position accuracy of such events.

\subsection{\texorpdfstring{$\delta$}{Delta}-electrons}
\label{subsect:delta-electrons}
\begin{figure}[!ht]
\centering
\begin{subfigure}[a]{0.5\textwidth}
\includegraphics[width=\textwidth]{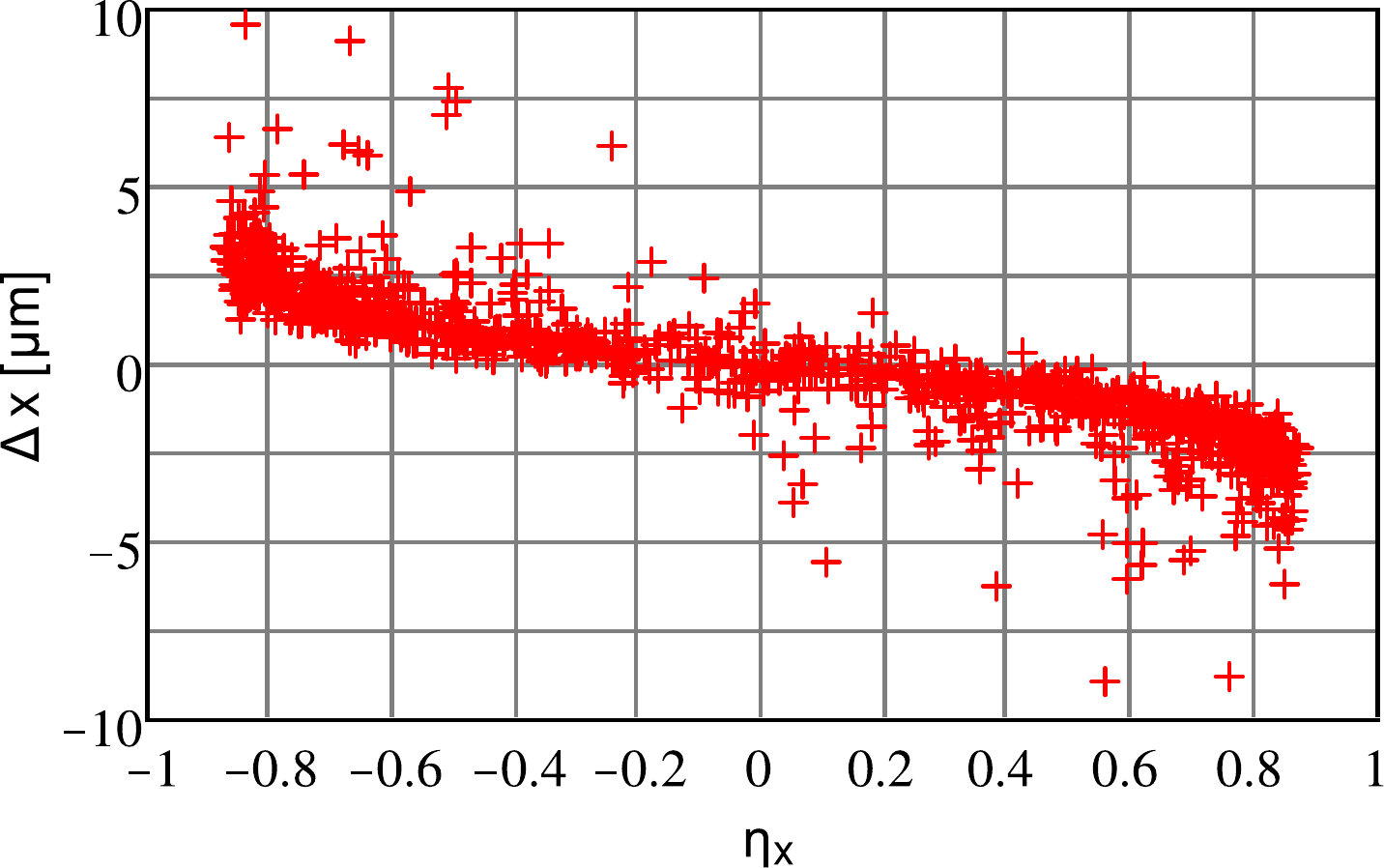}
\caption{ }
\label{fig:Dx-eta_medQ}
\end{subfigure}%
~
\begin{subfigure}[a]{0.5\textwidth}
\includegraphics[width=\textwidth]{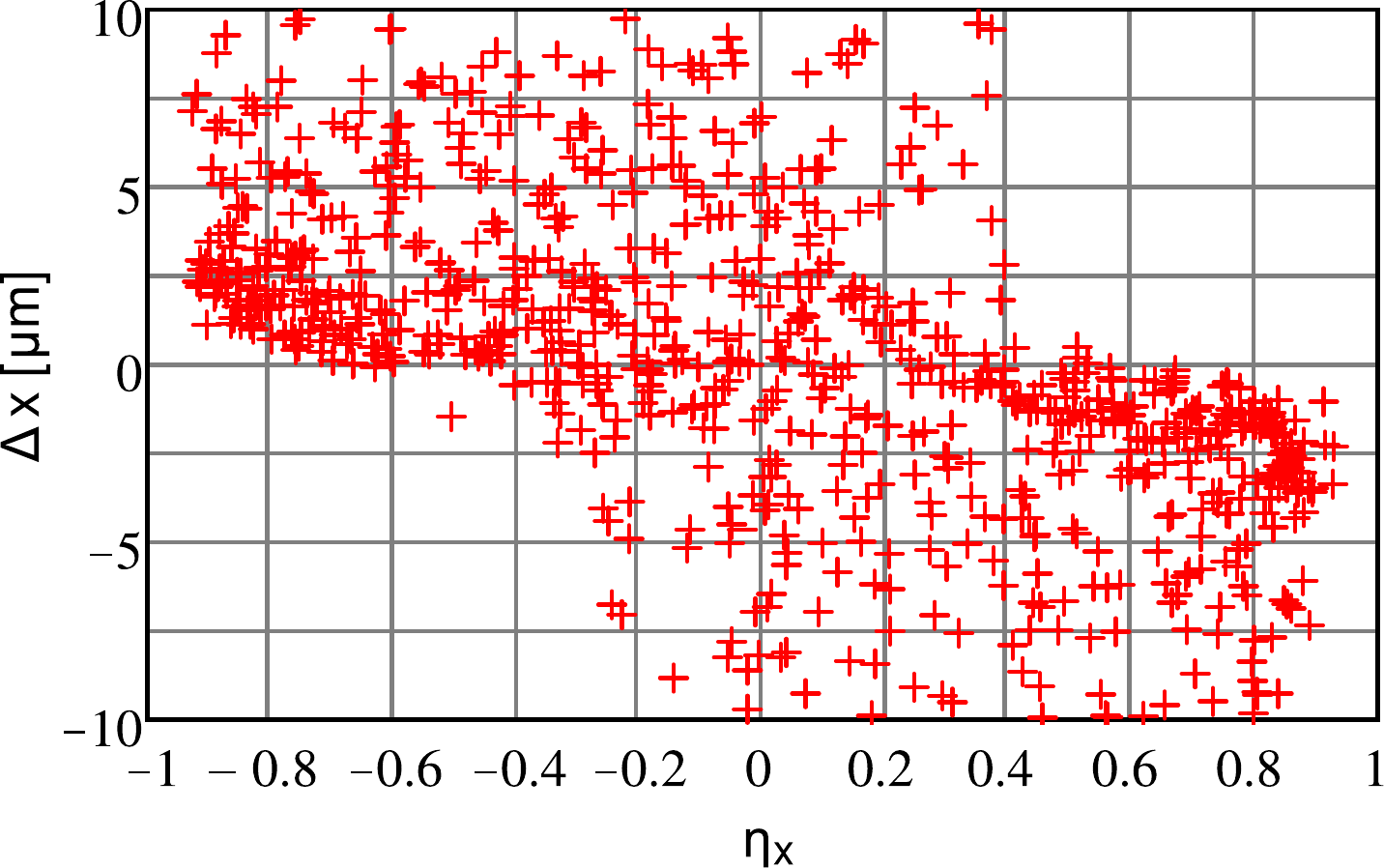}
\caption{ }
\label{fig:Dx-eta_hiQ}
\end{subfigure}%
\caption{Scatter plot of the residuals $\Delta x = x_\mathrm{DUT} - x_\mathit{true}$ versus $\eta_x$ for \emph{x-cls}\,$ = 2$,
(a) for cluster charges 16\,ke < \emph{Q} < 20\,ke, and
(b) for \emph{Q} > 20\,ke.
Events with $\delta$-electrons are responsible for the events at high $|\Delta x|$.}
\label{fig:Dx-eta_med-hiQ}
\end{figure}

The influence of energetic knock-on electrons ($\delta $-electrons) on the spatial resolution of cluster-size-two~events has been investigated by selecting events with cluster charges 16\,ke < \emph{Q} < 20\,ke and \emph{Q} > 20\,ke. The results for the $x$-direction are shown in Figure\,\ref{fig:Dx-eta_med-hiQ}; for the $y$-direction they are similar. 

The most probable value of the charge distribution is $Q \approx 11$\,ke. Figure\,\ref{fig:D-eta_XT0} shows that with the cut $Q < 16$\,ke there are hardly any events outside of a narrow band in the $\Delta x$ vs.\,$\eta _x$\,plot. In Figure\,\ref{fig:Dx-eta_medQ}, where events with 16\,ke < $Q$ < 20\,ke are selected, the fraction of events outside of the band is significantly higher, and for $Q$ > 20\,ke (Figure\,\ref{fig:Dx-eta_hiQ}) most events are outside, and the spatial resolution is seriously degraded. It is concluded that removing events with charges exceeding $\approx 1.5$\,times the most probable value, essentially removes the effects of energetic $\delta $-electrons.

\subsection{Incident angle}
\label{subsect:Incident angle}
\begin{figure}[!ht]
\centering
\begin{subfigure}[a]{0.5\textwidth}
\includegraphics[width=\textwidth]{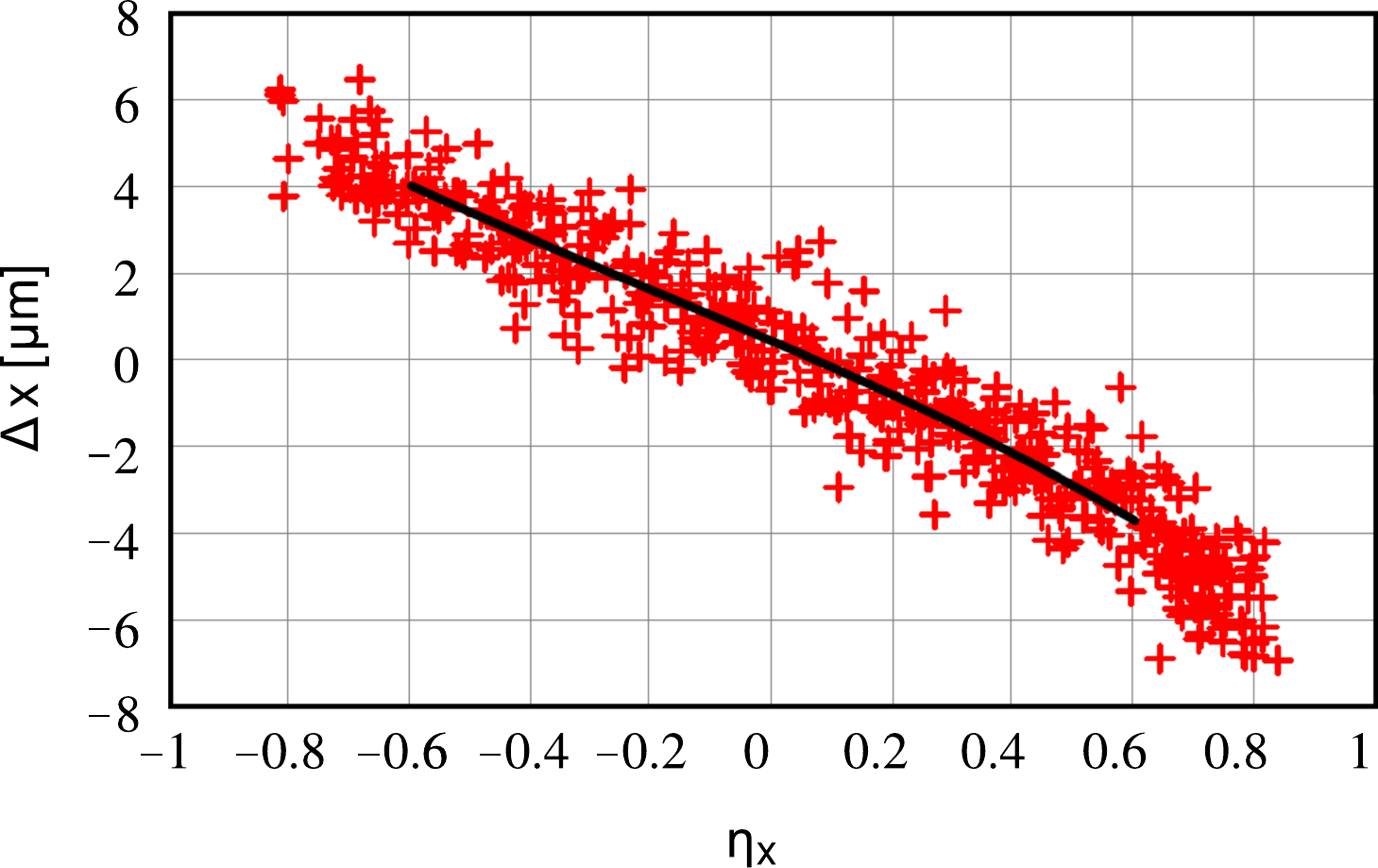}
\caption{ }
\label{fig:Dx-eta_small-angles}
\end{subfigure}%
~
\begin{subfigure}[a]{0.5\textwidth}
\includegraphics[width=\textwidth]{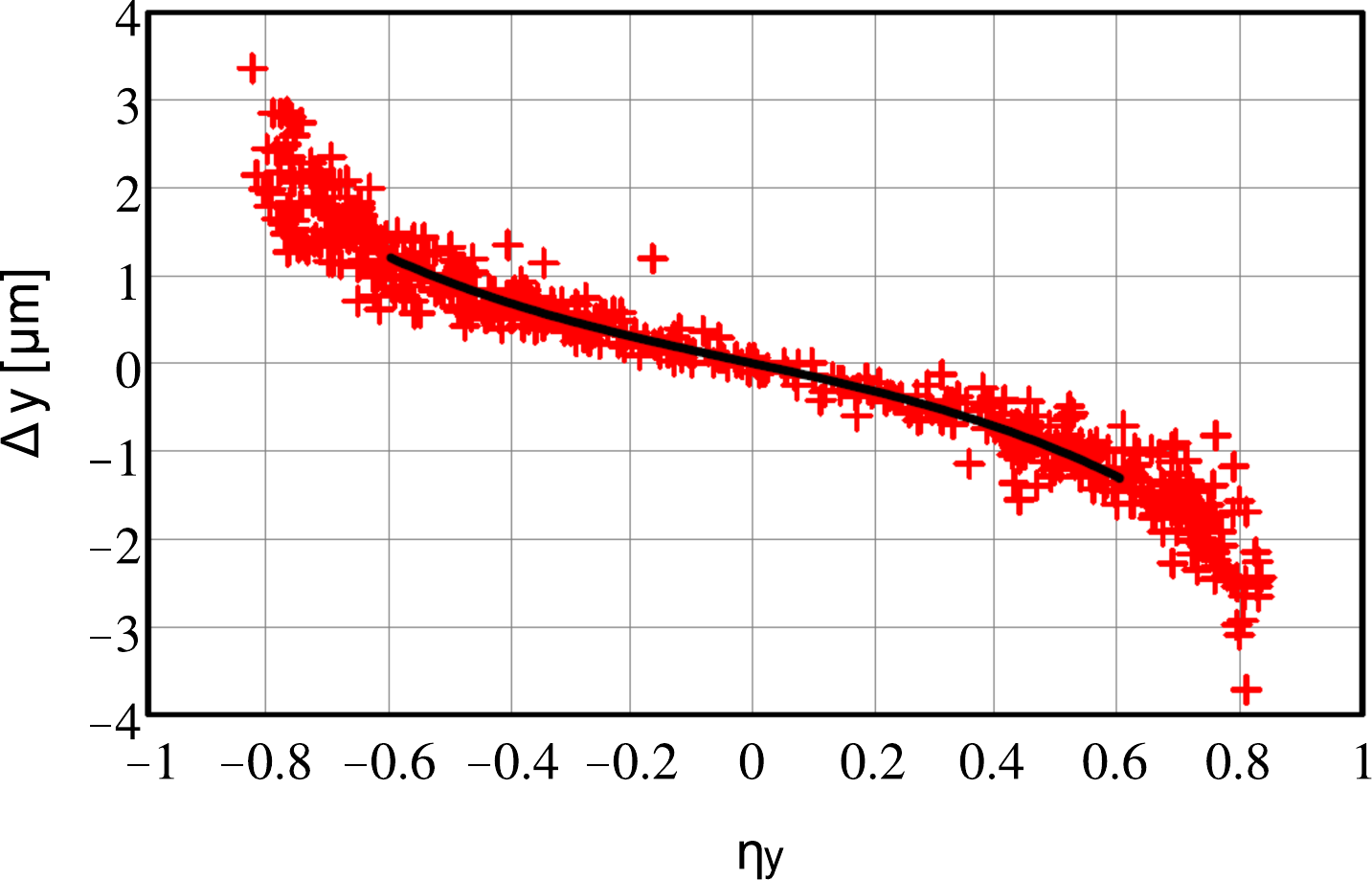}
\caption{ }
\label{fig:Dy-eta_small-angles}
\end{subfigure}%
\caption{(a) Scatter plot of the residuals $\Delta x = x_\mathrm{DUT} - x_\mathit{true}$ versus $\eta_x$, and (b) $\Delta y$ versus $\eta_y$ for an incident angle of \ang{5} relative to the normal in the \SI{25}{\um}\,direction (\emph{x}).
Note the factor 2 difference in vertical scale.
Whereas the $\Delta y$ distribution is not affected by the incident angle and is the same as in Figure\,\ref{fig:D-eta_XT0}, the $\Delta x$ is markedly different:
Both slope and width increase significantly.
The solid black lines are regression of third-order polynomials for $ -0.6 < \eta < + 0.6$, which can be used to correct $x_\mathrm{DUT}$ and $y_\mathrm{DUT}$.}
\label{fig:D-eta_small-angles}
\end{figure}

\begin{figure}[!ht]
\centering
\includegraphics[width=0.5\textwidth]{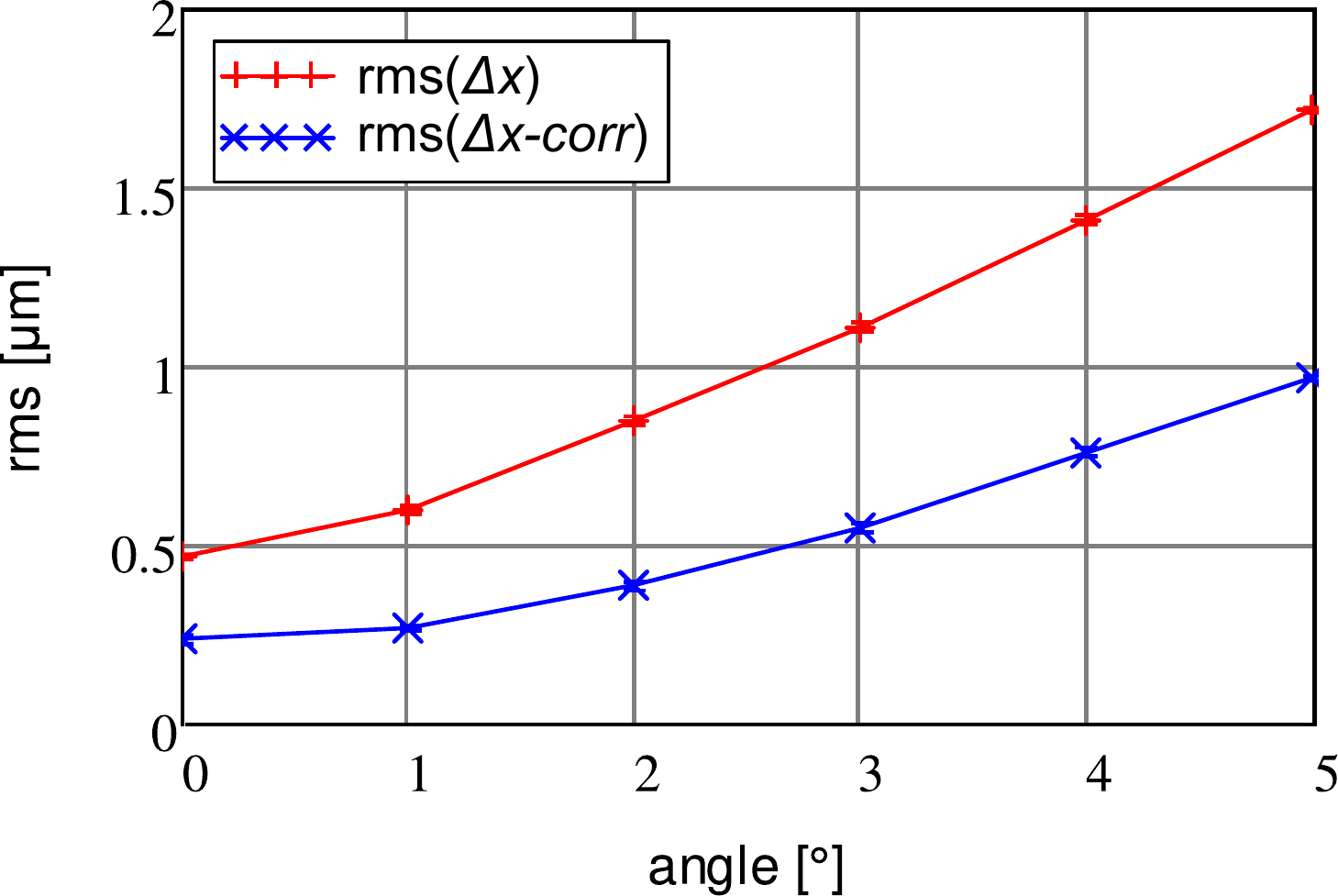}
\caption{rms of the residuals distributions $\Delta x = x_\mathrm{DUT} - x_\mathit{true}$ and $\Delta x\text{\small -corr} = x_\mathrm{DUT,corr} - x_\mathit{true}$ as a function of the incident angle relative to the \SI{25}{\um}\,direction (\emph{x}) for $|\eta _x| < 0.4$.
For \emph{rms}, the position of the pixel boundary of the \emph{x-cls}\,$=2$ cluster is assigned to $x_\mathrm{DUT}$, and for \emph{rms-corr}\,the third-order regression, which is shown as black line in Figure\,\ref{fig:Dx-eta_small-angles}, is subtracted.}
\label{fig:x-rms_small-angles}
\end{figure}

Next, the sensitivity of the method to deviations of up to \ang{5} from normal incidence of the beam is investigated. The study uses simulated events generated as described at the beginning of this section, with the only difference that the incident angle relative to the normal in the \SI{25}{\um}~direction was varied between \ang{0} and \ang{5} in \ang{1} steps.
Figure\,\ref{fig:D-eta_small-angles} shows for the \ang{5} simulations the scatter plots $\Delta x = x_\mathrm{DUT} - x_\mathit{true}$  and $\Delta y = y_\mathrm{DUT} - y _\mathit{true}$ versus $\eta$ for \emph{x-cls}\,$ = 2$ and \emph{y-cls}\,$ = 2$ events, respectively. Whereas the $\Delta y $\,distribution is the same as in Figure\,\ref{fig:D_XT0} and thus not affected by the \ang{5} angle, the $\Delta x $\,distribution is.
As expected three changes of the $\Delta x$\,distribution are observed:
 \begin{enumerate}
   \item the fraction \emph{x-cls}\,$ = 2$ events increases,
   \item the value of the slope | d$ \Delta x / \mathrm{d} \eta_x$ | increases, and
   \item the width of the $\Delta x $\,band increases.
 \end{enumerate}
The cause for item 1 is the increase of the length of the projection of the track on the readout plane which is proportional to the tangent of the incident angle.
Items 2 and 3 directly affect the $x$\,resolution.
This can be seen in Figure\,\ref{fig:x-rms_small-angles} which shows the dependence of the \emph{rms} of the $\Delta x$\,distribution for $| \eta _x| < 0.4$ on the angle of incidence. For the \emph{rms}, the position of the pixel boundary of the \emph{x-cls}\,$=2$ cluster is assigned to $x_\mathrm{DUT}$, and for \emph{rms-corr} the third-order regression, which is shown as black line in Figure\,\ref{fig:Dx-eta_small-angles}, is subtracted. It can be concluded that, as long as angles are below a few degrees, position resolutions for cluster-size-two~events of \SI{1}{\um} can be achieved by the proposed method.

\subsection{Summary}
\label{subsect:Summary}
A method is proposed to determine the track-position resolution of beam telescopes using silicon strip or pixel detectors with readout with charge digitization and normally incident beam. The method exploits the excellent spatial resolution of genuine clusters with projected cluster-size of two. It is straight-forward and easily programmed: cluster-size-two~events are selected, the charge asymmetry of the two signals is calculated, events in a given charge-asymmetry interval are selected, and the pixel (or strip) boundary of the two readout elements of the cluster is taken as reconstructed position. Position resolutions of about \SI{1}{\um} or less are achieved. The spatial resolution can be further improved to values below \SI{0.5}{\um} by a small correction for the shift of the reconstructed position relative to the pixel boundary, which can be obtained directly from the data as discussed in Section\,\ref{subsect:Simulation}. The main disadvantage of the method is that large event samples are required, as the fraction of events with projected genuine cluster-sizes of two is small; depending on the sensor pitch about 5 to 20\,\%. Cuts on the charge-asymmetry further reduce the sample by a factor two or more. The robustness of the method with respect to cross-talk, electronics noise, $\delta $-electrons and angular deviations from normal incidence of a few degrees has been demonstrated using simulated data. In Section\,\ref{sect:Results} the method will be used to measure the track-position resolution of the beam telescope of the DESY~II~Test~Beam~Facility.

In principle, the same method should be applicable for segmented silicon detectors with binary readout at normal incidence. In the absence of cross-talk and energetic $\delta $-electrons, the cluster-size-two~events will be confined to regions of widths of about $\Delta_{\mathit{cls} = 2} = f_{\mathit{cls} = 2} \cdot \mathit{pitch}$ centred at the boundaries between the electrodes. The fraction of events with cluster-size two is denoted $f_{\mathit{cls} = 2}$, and the pixel pitch \emph{pitch}. The \emph{rms} resolution will be approximately $\Delta_{\mathit{cls} = 2}/\sqrt{12}$. By changing the threshold, $\Delta_{\mathit{cls} = 2}$ can be changed and the resolution optimised. However, the method will be influenced by electronics noise, cross-talk and by energetic $\delta $-electrons, which should be investigated using simulated or experimental data.

\section{Validation using experimental data}
\label{sect:Validation}

\subsection{Experimental setup and data taking}
\label{sect:Setup}

\begin{figure}[!ht]
\centering
\includegraphics[width=0.8\textwidth]{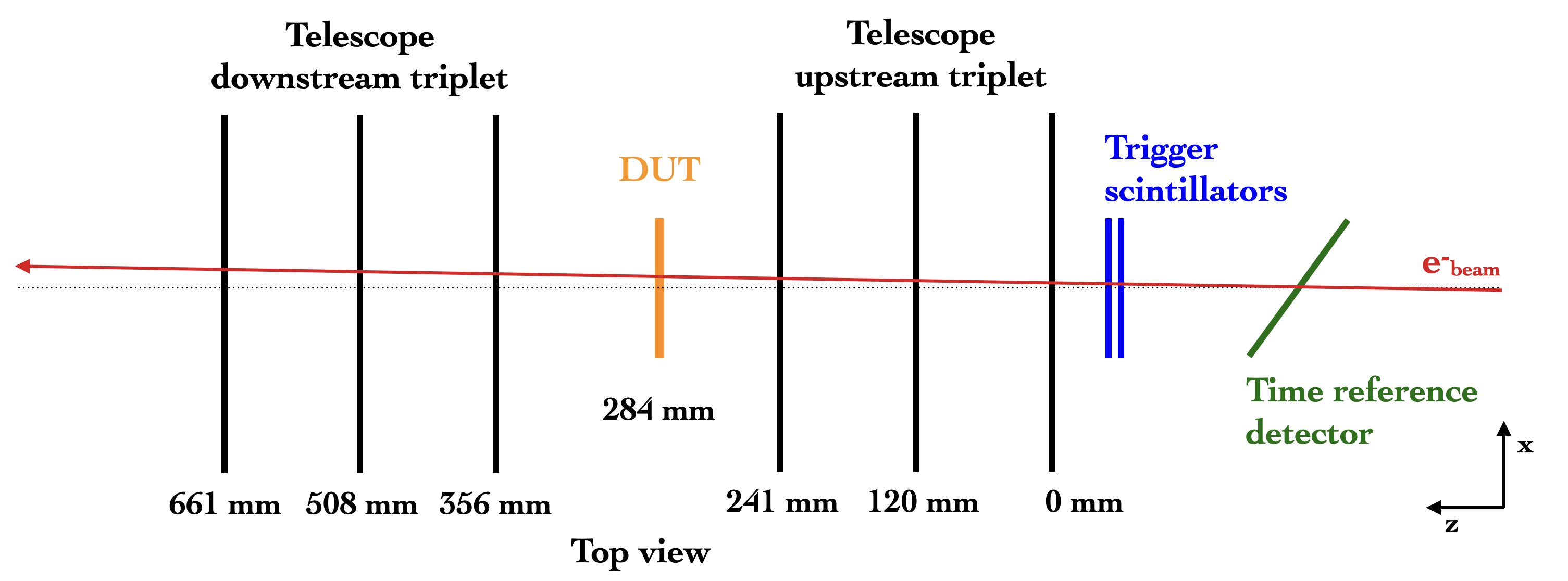}
\caption{Layout of the test beam setup: the six telescope planes, two trigger scintillators, the time reference detector, and the DUT.
The DUT material, relevant for multiple scattering and the energy loss of the electrons is:
$150 + \SI{150}{\um}$ Si for the sensor and the RD53A chip, respectively, the SnAg bumps, and a 1\,mm thick aluminum support plate on the downstream side of the sensor.}
\label{fig:Layout}
\end{figure}

In this section, the experimental setup in the DESY~II~Test~Beam~Facility, the sensor used and the data-taking conditions are briefly summarised. 
Figure\,\ref{fig:Layout} shows a sketch of the test beam setup\,\cite{Diener:2019}. A pair of plastic scintillators provides the trigger. The beam track-position at the DUT is measured by two beam telescope arms\,\cite{Jansen:2016}, one upstream and one downstream of the DUT. Each arm consists of 3 planes of Mimosa26 Monolithic Active Pixel Sensors (MAPS) with a pitch of \SI{18.4}{\um}$\,\times \, \SI{18.4}{\um}$ thinned down to a thickness of about \SI{50}{\um}. The readout is binary, and the single-plane spatial resolution is \SI{3.2}{\um}. The MAPS are read out in  \SI{115}{\us} frames and a pixel sensor (Time REF) upstream of the scintillators is used to select the track in coincidence with the signal in the DUT. Time REF is a CMS\,Phase-1 pixel detector\,\cite{Adam:2021} with a pixel size of \SI{100}{\um}$\,\times \, \SI{150}{\um}$ and 8-bit digital readout at a frequency of 40\,MHz. The DUT is mounted on computer-controlled stages, which allow moving the DUT horizontally and vertically, and rotating it around one axis.

The DUT is a CMS\,Phase-2 prototype pixel sensor\,\cite{Pixel:2023} read out by the RD53A\,chip. 
The sensor design is presented in Sec.\,\ref{subsect:Simulation}.
The linear front-end section of the RD53A chip\,\cite{Dimitrievska:2020}, bump-bonded to the sensor, is used for the readout. The chip is thinned down to a thickness of \SI{150}{\um}. The readout operates at a frequency of 40\,MHz, and using the Time-over-Threshold (ToT) method, digitises the charge above an adjustable threshold with 4-bit accuracy. The relation between the ToT value and the input charge can also be adjusted. A typical setting used for the measurements yielded a value of about 11 ToT units for a charge of 11\,ke. It is noted that the relation between charge and ToT is only approximately linear.

The data were taken at room temperature and at a bias voltage of 120\,V. The dark current of about \SI{3}{\uA} and the power dissipation of the RD53A chip was sufficiently low so that air-flow cooling was sufficient. The electronics noise was about 70\,e and the threshold for recording hits about 1250\,e. The track-position resolutions from the upstream ($\sigma _\mathit{up}$) and the downstream ($\sigma _\mathit{down}$) beam telescope arms extrapolated to the DUT are obtained using the "GBL Track Resolution Calculator" program of Ref.\,\cite{Spannagel:2016}, with the geometry and the materials given in Figure\,\ref{fig:Layout}.  The resolution values are reported in Table\,\ref{tab:RunPar} together with $\sigma _{0.5\,(\mathit{up\, + \, down})}=0.5\cdot\sqrt{\sigma_\mathit{up}^2+\sigma_\mathit{down}^2}$, which corresponds to the track-position resolution assuming weights of 0.5 for the upstream and the downstream beam-track positions when calculating the residuals with respect to the position measured by the DUT. Using the weights $\sigma_\mathit{up}/(\sigma_\mathit{up}+\sigma_\mathit{down})$ and $\sigma_\mathit{down}/(\sigma_\mathit{up}+\sigma_\mathit{down})$, a value of \SI{4.19}{\um} is obtained, and the general broken-line fit performed through all six beam telescope planes gives a value of \SI{3.94}{\um}.

\begin{table} [!ht]
\caption{Beam energy, track-position resolutions (upstream, downstream and mean value) extrapolated to the DUT, and data-taking conditions: bias voltage, temperature, electronic noise, and threshold.}
\label{tab:RunPar}
\centering
\begin{tabular}{c|c|c|c||c|c|c|c}
$E_e$ & $\sigma _\mathit{up}$ & $\sigma _\mathit{down}$ & $\sigma _{0.5\,(\mathit{up\, + \, down})} $ & $V_b$ & T & \emph{Noise} & \emph{Thr}   \\
\,[GeV] & [\unit{\um}] &  [\unit{\um}] &  [\unit{\um}] & [V] & [$^\circ$C] & [e] & [e]\\
\hline
5.2& 5.14 &7.01 & 4.35 & 120 & $\approx +20$ & $\approx 70$ & $\approx 1250$\\
\end{tabular}
\end{table}

It is noted that there are several differences between the actual measurements and the simulation.
Bond-pads, bond-balls, and the RD53A chip were not included in the simulation, and instead of using 5.2\,GeV electrons, 40\,GeV pions were simulated, as these data were readily available.
It is also observed that the purpose of the simulation is to illustrate the method and evaluate its robustness under various conditions, rather than to replicate the data presented in this section.

\subsection{Results}
\label{sect:Results}

In this section the data discussed in Section\,\ref{sect:Setup} are used to determine the position resolution of the tracks reconstructed in the beam telescope and extrapolated to the DUT. The resolution of the mean position of the upstream and the downstream beam telescope arms can be obtained in two ways: from half the difference of the position of the tracks reconstructed in the two arms and extrapolated to the DUT, and from the residuals of the positions from cluster-size-two~events and the mean beam track-position at the DUT. The comparison of the two results is a cross-check of the proposed method. In addition, the cluster-size-two method allows determining the resolutions of the upstream and the downstream beam telescope arms separately.

The reconstruction of the beam tracks and the alignment between DUT and beam tracks has been performed as described in Ref.\,\cite{Ebrahimi:2021}. 
Events were selected that have a distance of less than $\SI{300}{\um}$ between the position reconstructed in the DUT and the beam tracks extrapolated to $x_\mathit{beam}$, $y_\mathit{beam}$ on the DUT. It is noted that the angles of incidence of the beam were obtained from the data as a part of the alignment procedure: the angles of the normal of the DUT plane relative to the beam were changed until the slopes of $\langle \Delta x \rangle$ versus $x_\mathit{beam}$ and of $\langle\Delta y \rangle$ versus $y_\mathit{beam}$ were zero; $\Delta x $ and $\Delta y $ are the residuals of the DUT and of the beam track-positions for all cluster-sizes. The angles thus determined were $< \ang{1.2}$, with an uncertainty of about \ang{0.1}.

\begin{figure}[!ht]
\centering
\includegraphics[width=0.6\textwidth]{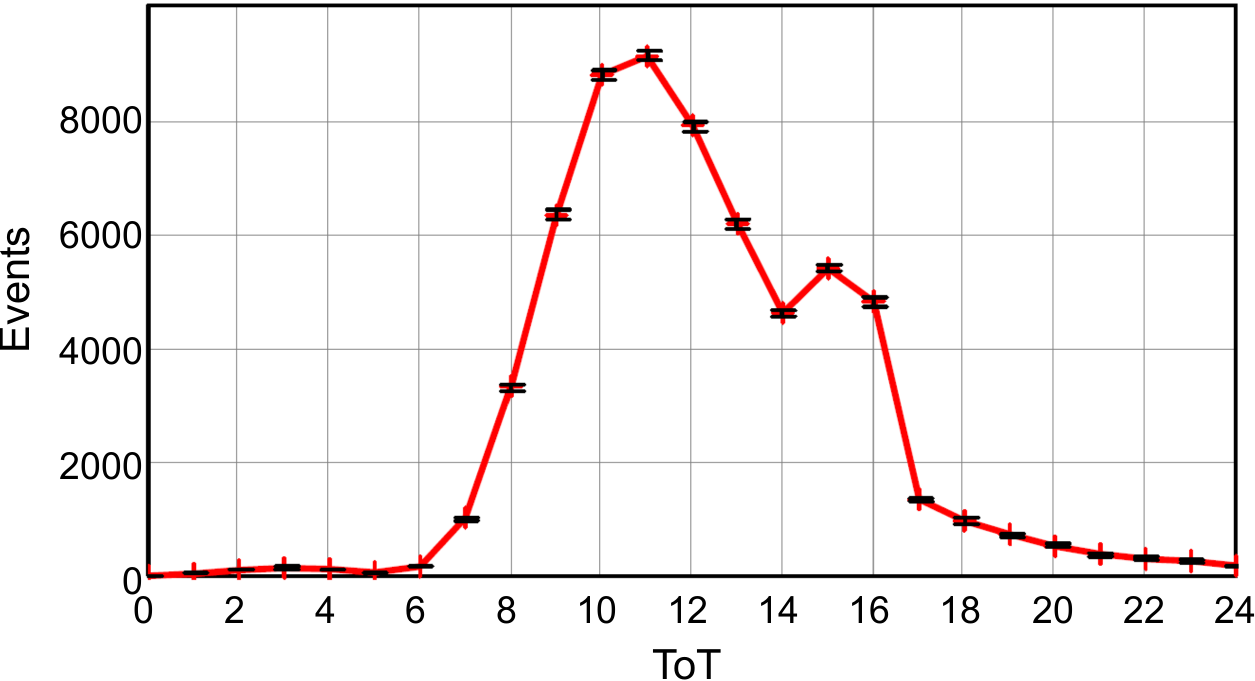}
\caption{Measured cluster charge spectrum for all events in ToT units.
The peak at 15 corresponds to single-hit events which saturate the 4-bit ToT-digitization of the RD53A chip. }
\label{fig:Qtot_non}
\end{figure}

Figure\,\ref{fig:Qtot_non} shows the measured charge spectrum in units of ToT. The peak at $Q = 15$\,ToT is due to cluster-size-one~events with a $Q$\,value which saturates the 4-bit range of the ToT. Up to ToT = 14, the spectrum is similar to the expected energy-loss distribution in \SI{150}{\um} silicon. However, there is a tail at lower \emph{Q}\,values, which is removed by the cut $Q > 6$\,{ToT}. Events with energetic $\delta $-electrons are removed by requiring $Q < 16$\,{ToT}.

\begin{figure}[!ht]
\centering
\begin{subfigure}[a]{0.5\textwidth}
\includegraphics[width=\textwidth]{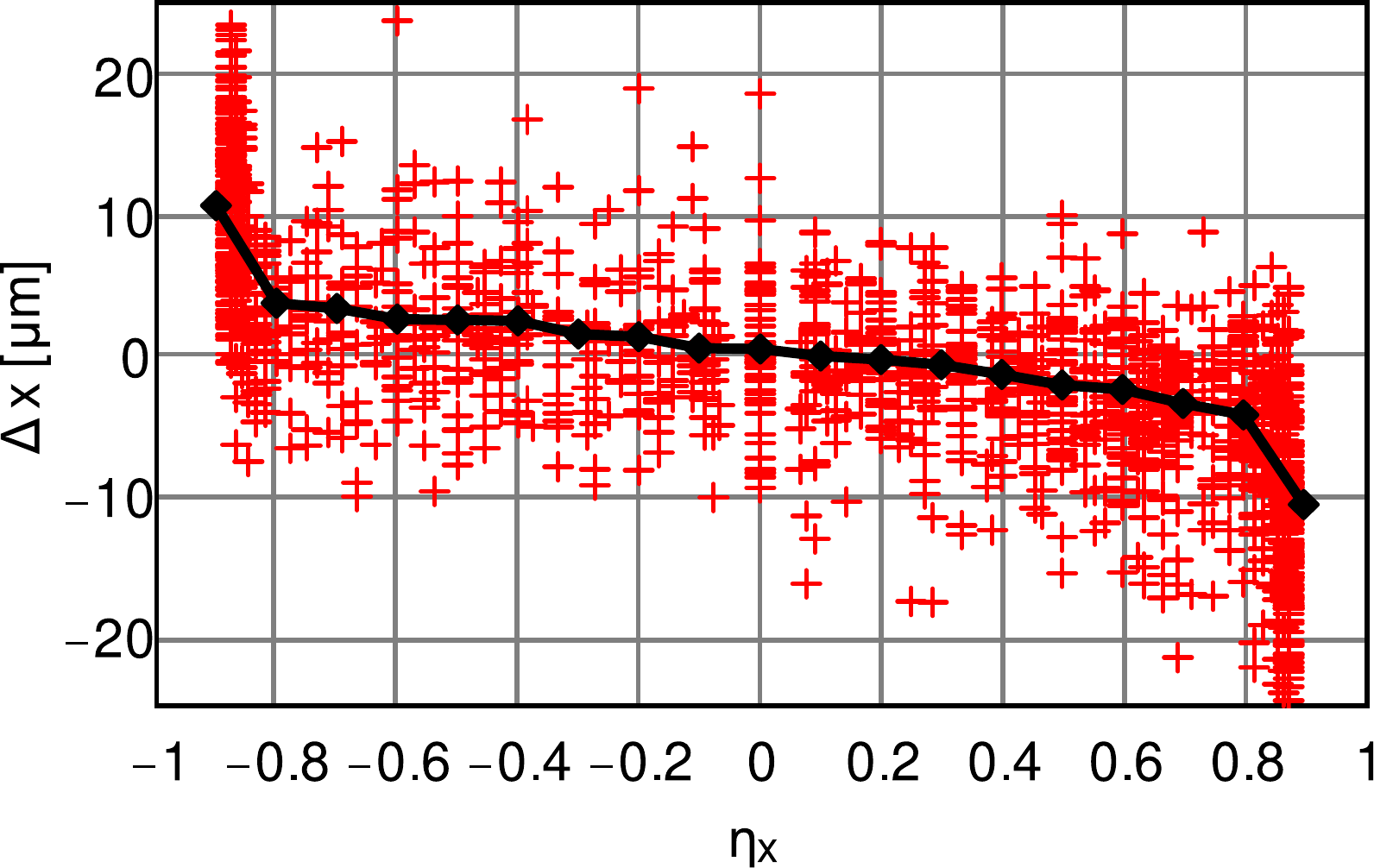}
\caption{ }
\label{fig:Dx-eta_non}
\end{subfigure}%
~
\begin{subfigure}[a]{0.5\textwidth}
\includegraphics[width=\textwidth]{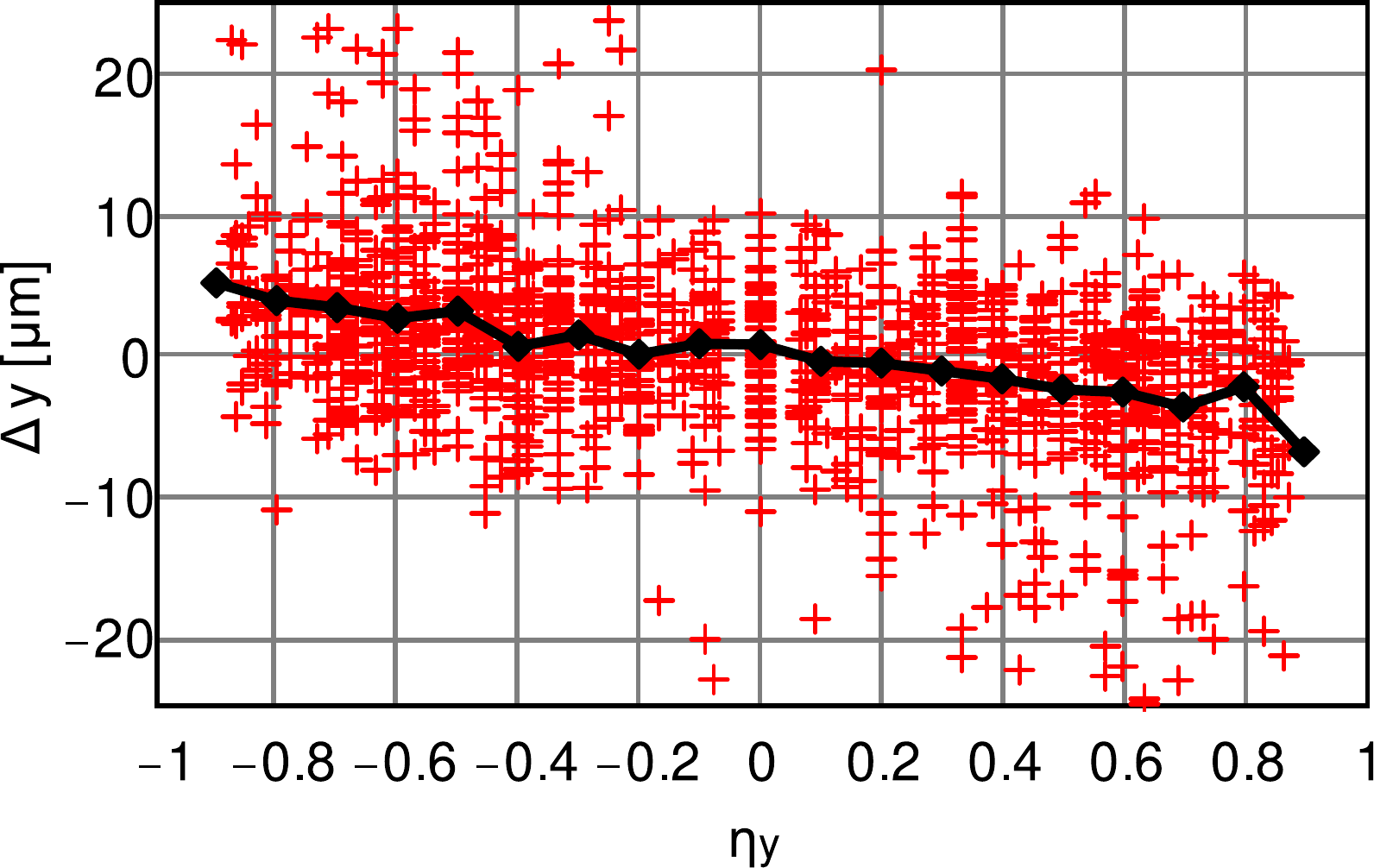}
\caption{ }
\label{fig:Dy-eta_non}
\end{subfigure}%
\caption{(a) Scatter plot of the residuals $\Delta x = x_\mathrm{DUT} - x_\mathit{beam}$ versus $\eta_x$, and (b) $\Delta y$ versus $\eta_y$ for 2000 events.
For $x_\mathit{beam}$ the mean of $x_\mathit{up}$ and $x_\mathit{down}$ at the position of the DUT is used and similarly for the $y$-direction.
The black diamond-shaped points are the median values of the $\Delta $\,distributions for 0.1\,intervals in $\eta $.
Their $\eta$\,dependencies are similar to the mean values shown in Figure\,\ref{fig:D-eta_XT0}, where the dependence is discussed.}
\label{fig:D-eta_non}
\end{figure}

Figure\,\ref{fig:Dx-eta_non} shows the scatter plot $\Delta x = x_\mathrm{DUT} -  x_\mathit{beam}$ versus $\eta_x$ for \emph{x-cls}\,$ = 2$ events; $x_\mathrm{DUT}$ is the $x$\,position of the pixel boundary of the cluster, and $x_\mathit{beam}$ the mean of the extrapolated $x$\,positions of the upstream ($x_\mathit{up}$) and downstream ($x_\mathit{down}$) beam telescope arms. The corresponding scatter plot for the $y$-direction is displayed in Figure\,\ref{fig:Dy-eta_non}. The black diamond-shaped points are the median values for the different $\eta $ bins. Different to Section\,\ref{sect:Method}, where the mean is used, the medians are shown, as they are less sensitive to outliers, which are seen in the scatter plots. To the left and to the right of the events at $\eta = 0$, there are empty bands, which result from the $Q$\,cut and the 4-bit ToT accuracy. Up to $|\eta| = 0.8$, the median values are approximately proportional to $ -\, \eta$, with values of $\pm\,\SI{4}{\um}$ at $\eta = \mp\,\num{0.8}$. In Figure\,\ref{fig:Dx-eta_non} many events with large $|\Delta x|$ are observed at $ |\eta_x | > 0.8$, which are caused by the cross-talk from genuine \emph{x-cls}\,$=1$ events. These events are absent in the $y$-direction where hardly any cross-talk is expected.

\begin{figure}[!ht]
\centering
\begin{subfigure}[a]{0.5\textwidth}
\includegraphics[width=\textwidth]{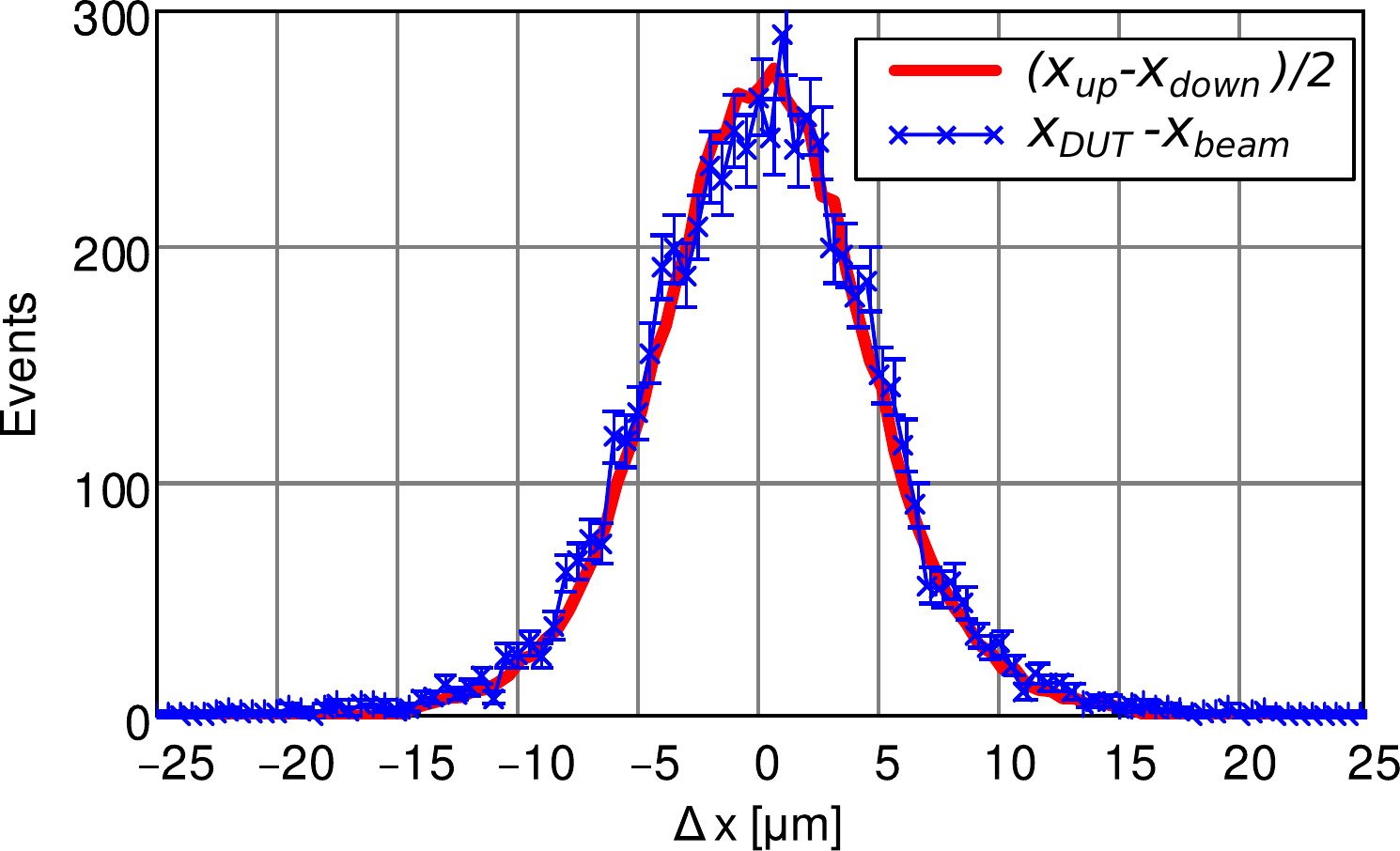}
\caption{ }
\label{fig:xDUT-beam_non}
\end{subfigure}%
~
\begin{subfigure}[a]{0.495\textwidth}
\includegraphics[width=\textwidth]{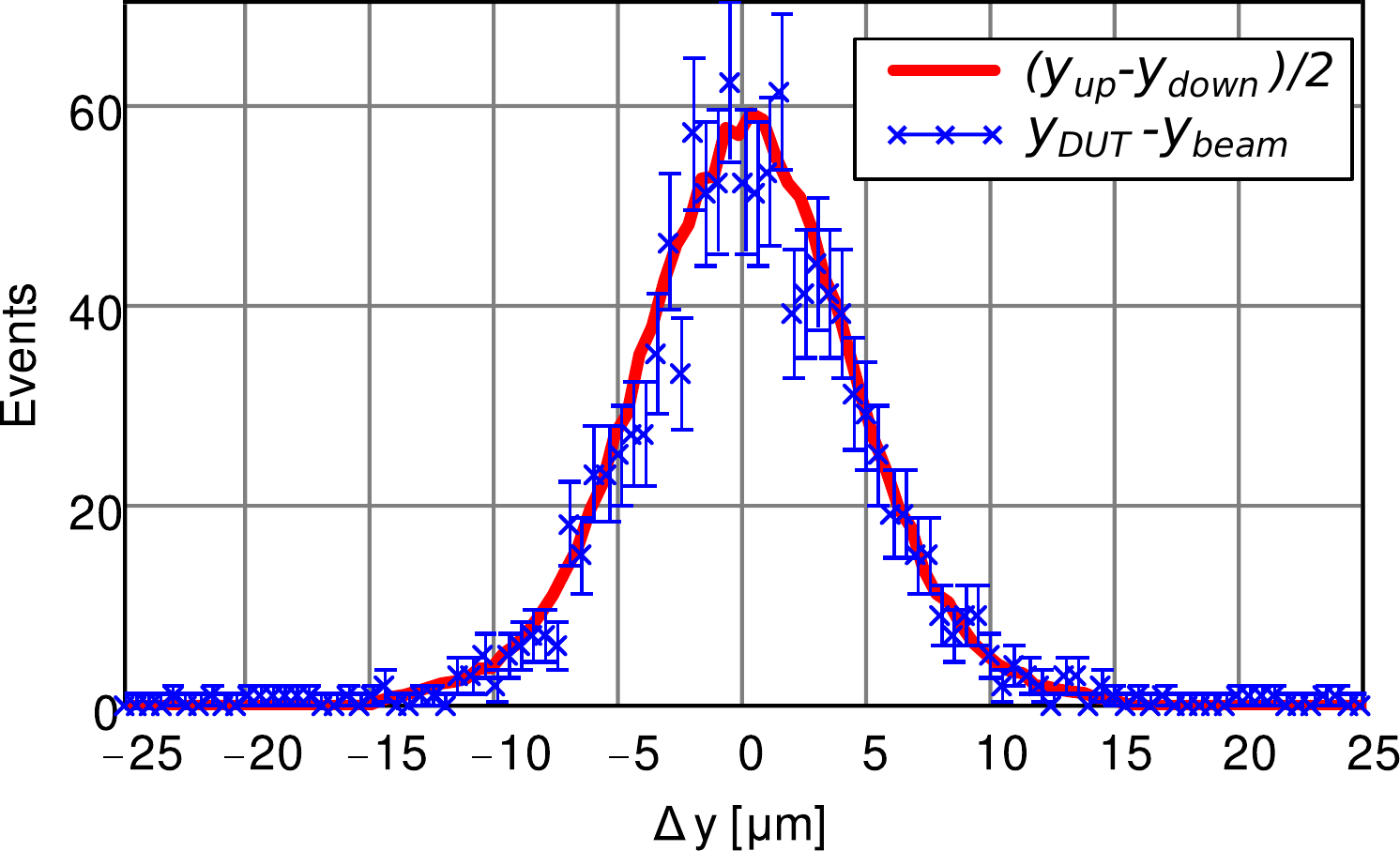}
\caption{ }
\label{fig:yDUT-beam_non}
\end{subfigure}%
\caption{(a) Comparison of the residual distribution $\Delta x = x_\mathrm{DUT} - x_\mathit{beam}$ for $|\eta _x | < 0.4$ (points with error bars) to the distribution of $(x_\mathit{up} - x_\mathit{down})/2 $ (continuous line).
For $x_\mathit{beam}$ the mean of $x_\mathit{up}$ and $x_\mathit{down}$ at the position of the DUT is used.
The $0.5 \cdot(x_\mathit{up} - x_\mathit{down})$ distribution is normalised to the $\Delta x$\,distribution.
(b) Same for the $y$-direction. }
\label{fig:DUT-beam_non}
\end{figure}

\begin{figure}[!ht]
\centering
\begin{subfigure}[a]{0.48\textwidth}
\includegraphics[width=\textwidth]{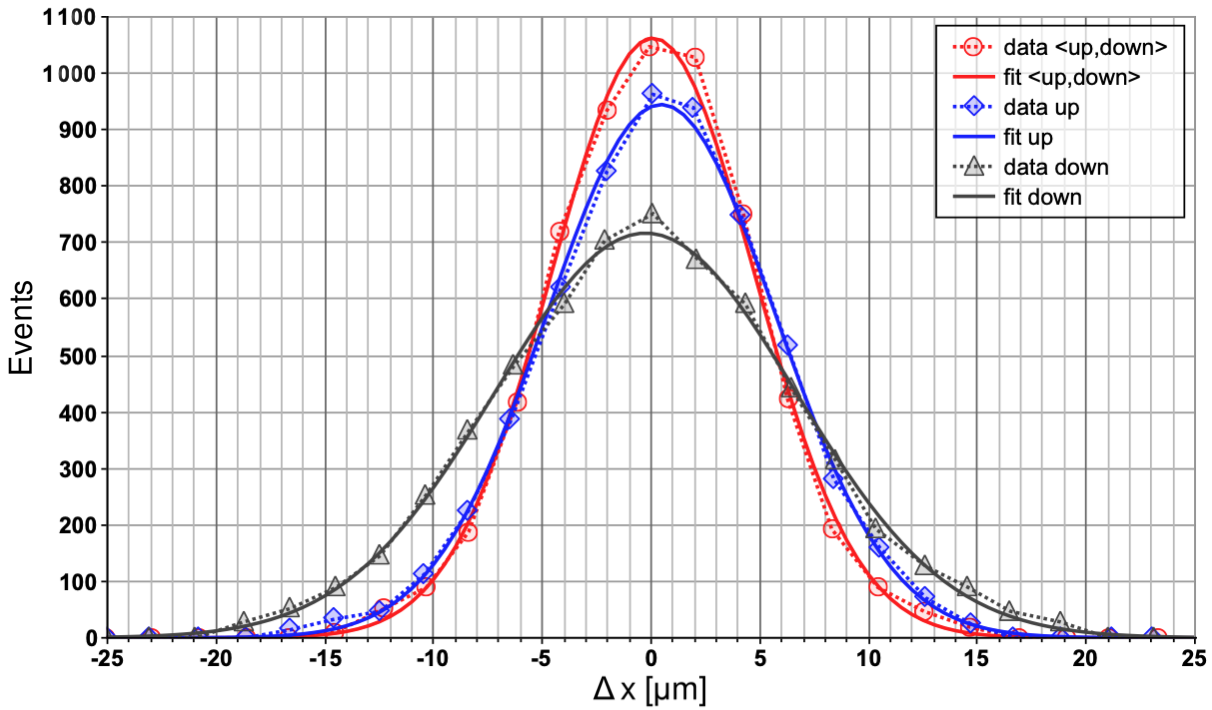}
\caption{ }
\label{fig:Dx-mud_non}
\end{subfigure}%
~
\begin{subfigure}[a]{0.48\textwidth}
\includegraphics[width=\textwidth]{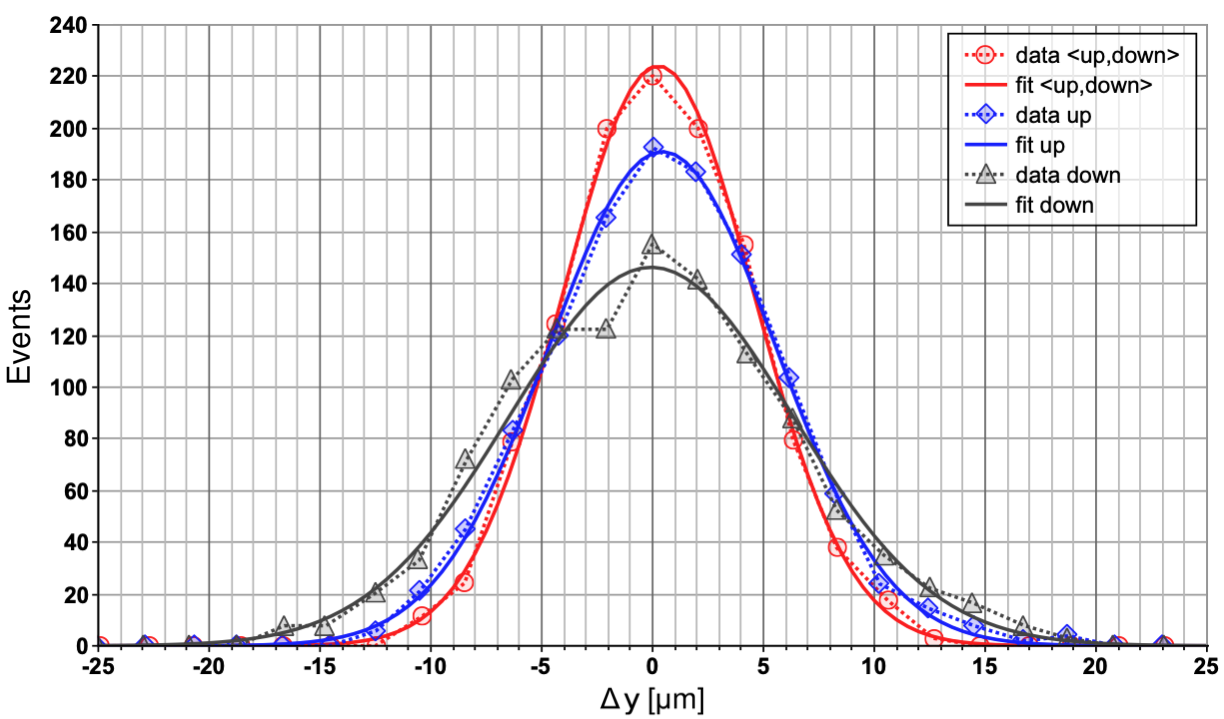}
\caption{ }
\label{fig:Dy-mud_non}
\end{subfigure}%
\caption{(a) Comparison of the residual distributions $x_\mathrm{DUT} - x_\mathit{beam}$, of $x_\mathrm{DUT} - x_\mathit{up}$ and of $x_\mathrm{DUT} - x_\mathit{down}$, for $|\eta _x | < 0.4$.
(b) Same for the $y$-direction.}
\label{fig:D-mud_non}
\end{figure}

In the following, the pixel boundaries of the two pixels of the clusters are assigned to $ x_\mathrm{DUT}$ and $y_\mathrm{DUT}$ and the small correction discussed in Section\,\ref{sect:Method} is not applied. Figure\,\ref{fig:DUT-beam_non} compares the residual distributions $\Delta x$ and $\Delta y$ of cluster-size-two~events with the distributions of  $0.5 \cdot (x_\mathit{up} - x_\mathit{down})$ and of $0.5 \cdot(y_\mathit{up} - y_\mathit{down})$ of all events for the $x$- and $y$-directions, respectively, with $|\eta | < 0.4$ and normalized to the $\Delta$ distributions. 
The agreement between the two distributions demonstrates that the track-position resolution can be determined with the proposed method.
From Table\,\ref{tab:RunPar} it can be seen that the calculated track-position resolutions at the DUT of the upstream and the downstream beam telescope arms differ. This results from the different distances between $z_\mathrm{DUT}$ and the adjacent planes of the two beam telescope arms (see Figure\,\ref{fig:Layout}), and is confirmed by Figure\,\ref{fig:Dx-mud_non}, which shows for the $x$-direction the distributions of the residuals
$x_\mathrm{DUT} - x_\mathit{beam}$, $x_\mathrm{DUT} - x_\mathit{up}$ and $x_\mathrm{DUT} - x_\mathit{down}$.
Similarly, Figure\,\ref{fig:Dy-mud_non} displays the distributions for the $y$-direction.

\begin{figure}[!ht]
\centering
\begin{subfigure}[a]{0.5\textwidth}
\includegraphics[width=\textwidth]{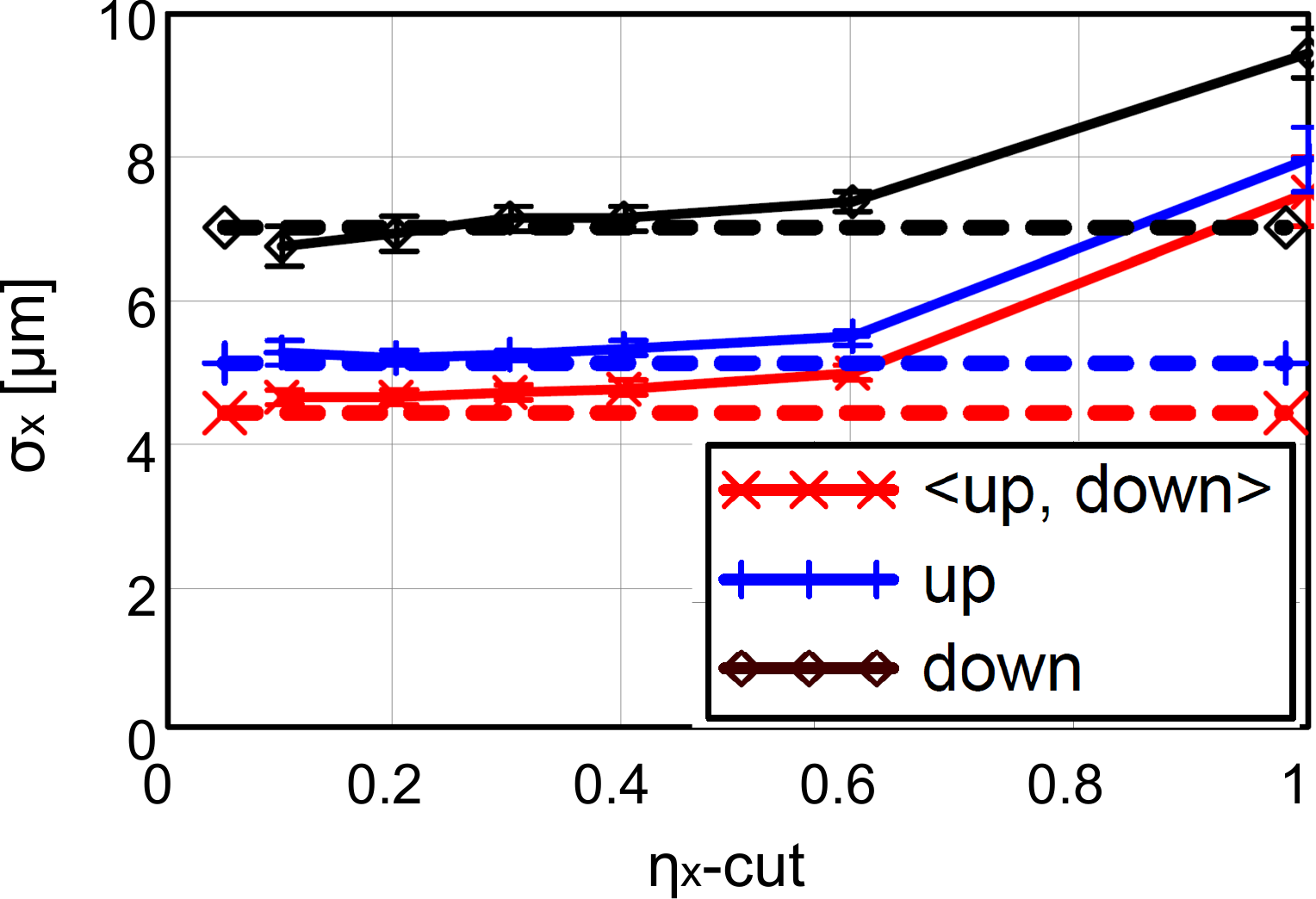}
\caption{ }
\label{fig:RMSx-eta_non}
\end{subfigure}%
~
\begin{subfigure}[a]{0.5\textwidth}
\includegraphics[width=\textwidth]{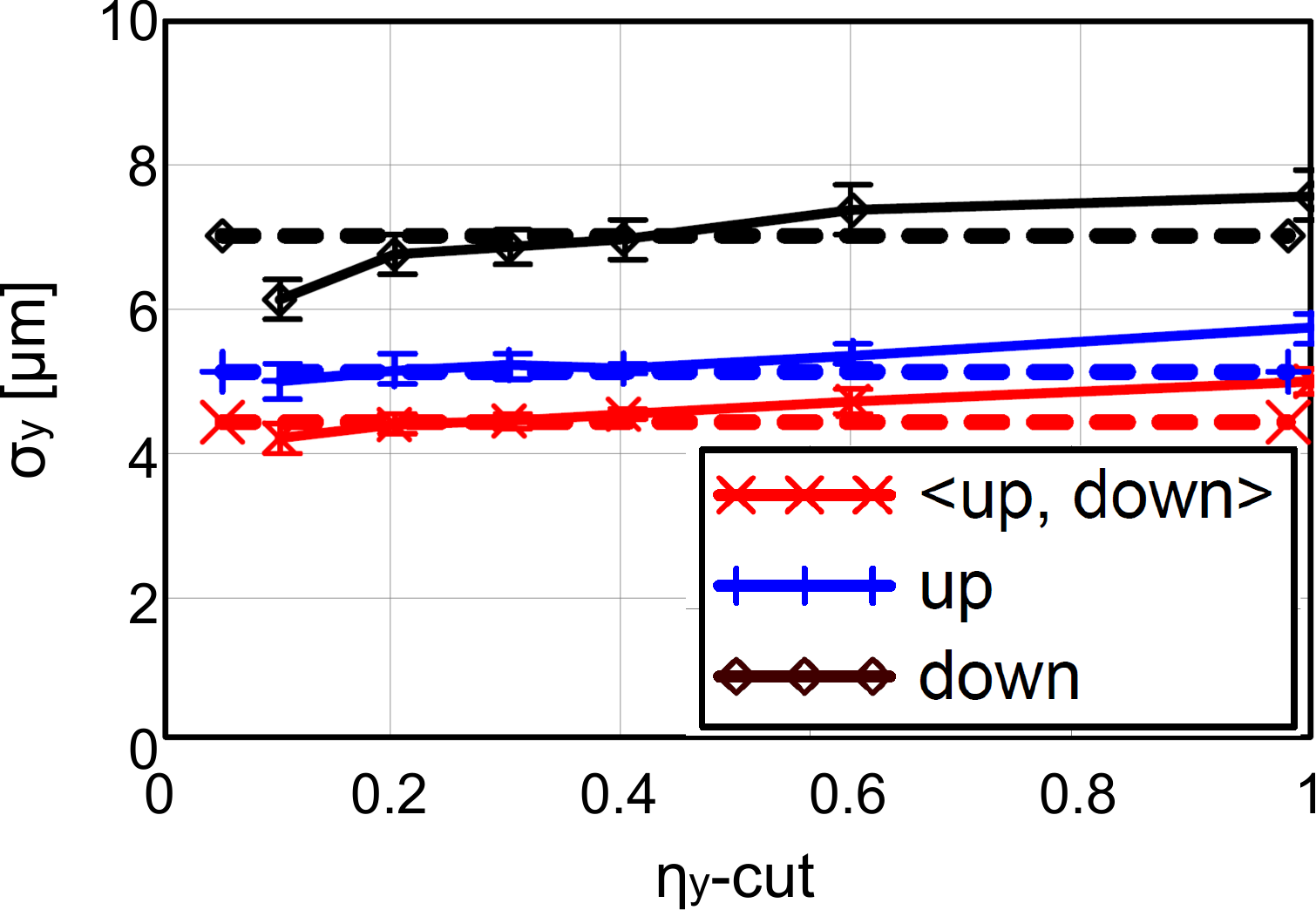}
\caption{ }
\label{fig:RMSy-eta_non}
\end{subfigure}%
\caption{(a) Resolutions, $\sigma $, obtained from Gaussian fits to the residual distributions $x_\mathrm{DUT} - x_\mathit{beam}$, of $x_\mathrm{DUT} - x_\mathit{up}$ and of $x_\mathrm{DUT} - x_\mathit{down}$ as a function of the $|\eta_x|$\,cut. (b) Same for the $y$-direction.
The dashed horizontal lines indicate the resolutions given in Table\,\ref{tab:RunPar}.}
\label{fig:RMS-eta_non}
\end{figure}

Figure\,\ref{fig:RMS-eta_non} shows the $\sigma $\,values obtained from Gaussian fits to the same residual distributions, as a function of the $|\eta|$\,cut together with the calculated track-position resolutions discussed in Section\,\ref{sect:Setup}. It can be seen that up to $|\eta| \lesssim 0.6$ the observed and the calculated $\sigma $\,values agree. The large increase in $\sigma_x$ towards $|\eta_x| = 1$ is caused by cross-talk. This increase is absent for $\sigma_y$, confirming the absence of significant cross-talk in the \SI{100}{\um} pitch direction.

\begin{table} [!ht]
\caption{Comparison of the calculated and the measured beam track-position resolutions. The values measured using the DUT (columns 2 to 4) are for $|\eta| < 0.4$.
The values of column 5 are obtained from the position difference of the upstream and the downstream beam telescope arms. For all values, statistical errors are presented, except for the first row, where uncertainties arise from systematic errors calculated by considering an error in the alignment procedure of $z_\mathrm{DUT}$ of $\pm$\SI{1}{mm}.}
\label{tab:Sig_non}
\centering
\begin{tabular}{c||c|c|c||c}
& $\sigma _\mathit{up}$\,[\unit{\um}] & $ \sigma _\mathit{down} $\,[\unit{\um}] & $\sigma _\mathit{0.5\,(up + down)} $\,[\unit{\um}] & $\sigma _\mathit{0.5\,(up - down)}^\mathit{beam} $\,[\unit{\um}]   \\
\hline
calculated & $5.14 \pm 0.06 $ & $7.01 \pm 0.07 $ & $4.35 \pm 0.05 $ & -- \\
measured in $x$ & $5.2 \pm 0.1$ & $7.0 \pm 0.2$ & $4.6 \pm 0.1$ & $4.4 \pm 0.04$  \\
measured in $y$ & $5.0 \pm 0.2$ & $6.6 \pm 0.3$ & $4.3 \pm 0.2 $& $4.4 \pm 0.12$  \\
\end{tabular}
\end{table}

In columns 2 to 4 of Table\,\ref{tab:Sig_non}, the calculated and the measured track-position resolutions for an $|\eta|$\,cut of 0.4 are reported with their uncertainties. The last column shows the $\sigma^{beam}_{0.5(up-down)} $\,values of one half of the residuals of the upstream and downstream beam track-positions, which measure (without information from the DUT) the resolution of the average of the upstream and the downstream beam track-positions extrapolated to the DUT. The agreement with the corresponding calculated value, $\sigma _\mathit{0.5\,(up + down)} $, demonstrates the validity of the beam-resolution calculations. The agreement between $\sigma _\mathit{0.5\,(up - down)}^\mathit{beam}$ measured using the beam telescope arms and $\sigma _\mathit{0.5\,(up + down)}$ measured using the DUT, demonstrates the validity of the proposed method. Similar agreements are observed between calculated and measured values of $\sigma _\mathit{up}$ and of $\sigma _\mathit{down}$ for both $x$- and $y$-directions. It is noted that an experimental method to determine  the track-position resolution of the upstream and the downstream beam telescope arms of the DESY~II~Test~Beam~Facility independently is not yet available.
It is concluded that cluster-size-two~events have sufficient accuracy to determine the position resolution of tracks reconstructed in the beam telescope and extrapolated to the DUT.

\section{Summary and conclusions}
\label{sect:Conclusions}

Beam tests are a standard technique to measure the spatial resolution of detectors. The residuals between the DUT and the beam track-position at the DUT are measured, which requires the precise knowledge of the beam track-position at the DUT. A straight-forward method is proposed to show how this can be achieved with a segmented silicon detector with readout with charge digitization. It relies on the well-known fact that the spatial resolution of segmented silicon detectors for tracks with normal incidence is excellent if they traverse the sensor close to the boundaries between the electrodes. In the proposed method, events with projected cluster-size-two are selected, the asymmetry, $\eta $, of the two charge values of the cluster is  calculated, and the boundary between the two elements is assigned to the position reconstructed in the DUT. Using simulated events for a sensor of \SI{150}{\um} thickness and \SI{25}{\um} $\times \,\SI{100}{\um}$ pixels, with a cut on the charge of less than 1.5\,times the most-probable charge, a spatial resolution below \SI{0.5}{\um} is found for $|\eta | \leq 0.4$. The same accuracy is obtained for the \SI{25}{\um} and the \SI{100}{\um}\,direction. Using the simulated events, the effects of cross-talk between pixels, electronics noise,  energetic $\delta $-electrons, and deviations from normal incidence up to \ang{5} are investigated, and it is found that with cuts on $\eta $ and on the total cluster charge, these can be controlled.

The method is applied to data taken with a pixel sensor as DUT in the  DESY~II~Test~Beam~Facility, which is equipped with two beam telescope arms, one upstream and one downstream of the DUT. For the DUT a CMS\,Phase-2 prototype pixel sensor with pixels of $\SI{25}{\um}\, \times \SI{100}{\um}$ is used. One method of determining the position resolution of beam tracks, which does not use the information from the DUT, uses the difference of the positions reconstructed by the two beam telescope arms extrapolated to the DUT; the other uses the difference of the mean position reconstructed by the two arms and the position reconstructed by the DUT using cluster-size-two~events. The results agree, which demonstrates the validity of the proposed method. This method also allows to determine experimentally the track-position resolution separately for the upstream and the downstream beam telescope arms, which was not possible so far. It is observed that the track-position resolutions of the upstream and downstream arms extrapolated to the DUT differ, which is confirmed by calculations. It is noted that the sensors are read out by the RD53A test chip, with a 4-bit accuracy of the charge measurement. Thus, the method also works for this coarse charge measurement. In principle, the method can also be applied to detectors with binary readout; however, further studies are needed to determine the limitations in this case.

Using data, which are not shown in this paper, with normal incidence in one view and shallow incidence of \ang{78.7} to the normal to the sensor plane in the other view, it can be shown that the spatial resolution for the view with normal incidence is not affected. If the track-position resolution of the beam telescope is the same in both views, which is normally the case, the proposed method allows to determine it for every data set even in angular scans. 
This method is simple, can be easily implemented, and enables a precise determination of the position resolution of segmented silicon detectors in test beams.

\section*{Acknowledgements}
\label{sect:Acknowledgement}

This work was supported by the German Federal Ministry of Education and Research (BMBF) in the framework of the "FIS-Projekt - Fortf\"uhrung des CMS-Experiments zum Einsatz am HL-LHC: Verbesserung des Spurdetektors f\"ur das Phase-2 Upgrade des CMS-Experiments".
We would like to thank the Tracker Group of the CMS Collaboration for providing the samples for this study.
The measurements leading to these results have been performed at the Test~Beam~Facility at DESY Hamburg (Germany), a member of the Helmholtz Association (HGF). EG, JS, GS and AV acknowledge the support by the Deutsche Forschungsgemeinschaft (DFG, German Research Foundation) under Germany's Excellence Strategy - EXC 2121 Quantum Universe - 390833306.


\appendix
\setcounter{figure}{0}
\setcounter{table}{0}

\section{Comparison of the cluster-size 1 and 2 methods}
\label{app:Comparison}

In this section the proposed method, called $\mathit{cls} = 2$ method, is compared to the method used in Ref.\,\cite{Terzo:2015, Koppenhoefer:2022, Ziemons:2022} which uses  cluster-size-one~events, and is called $\mathit{cls} = 1$ method in the following.

  \begin{figure}[!ht]
   \centering
    \includegraphics[width=0.5\textwidth]{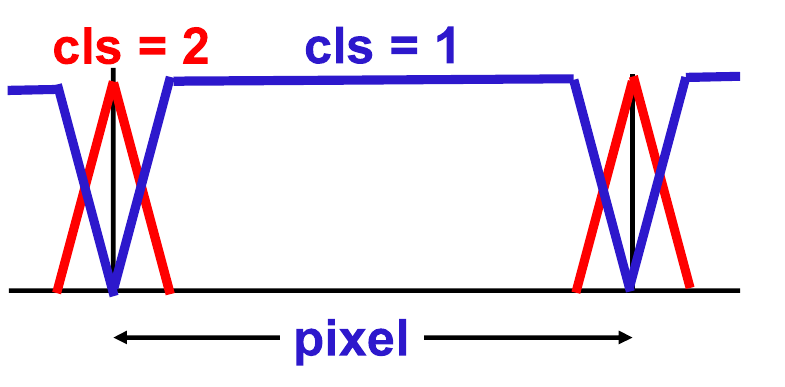}
   \caption{Schematic comparison of the spatial distributions of the $\mathit{cls} = 1$ and of the $\mathit{cls} = 2$ events.}
  \label{fig:cls12}
 \end{figure}

Figure\,\ref{fig:cls12} provides a schematic explanation of the two methods. The spatial distributions of $\mathit{cls} = 1$ and $\mathit{cls} = 2$ events for normally incident particles uniformly distributed over a silicon pixel sensor are shown. The $\mathit{cls} = 2$ events occupy narrow regions around the pixel boundaries, and the $\mathit{cls} = 1$ events the remainder. The $\mathit{cls} = 1$ events are reconstructed in the centre of the pixels, resulting in a box-type distribution for the residuals of reconstructed and true positions. After convolution with the track-position resolution, the distribution can be described by the difference of two error functions with width parameters corresponding to the track-position resolution.

\begin{figure}[!ht]
   \centering
   \begin{subfigure}[a]{0.5\textwidth}
    \includegraphics[width=\textwidth]{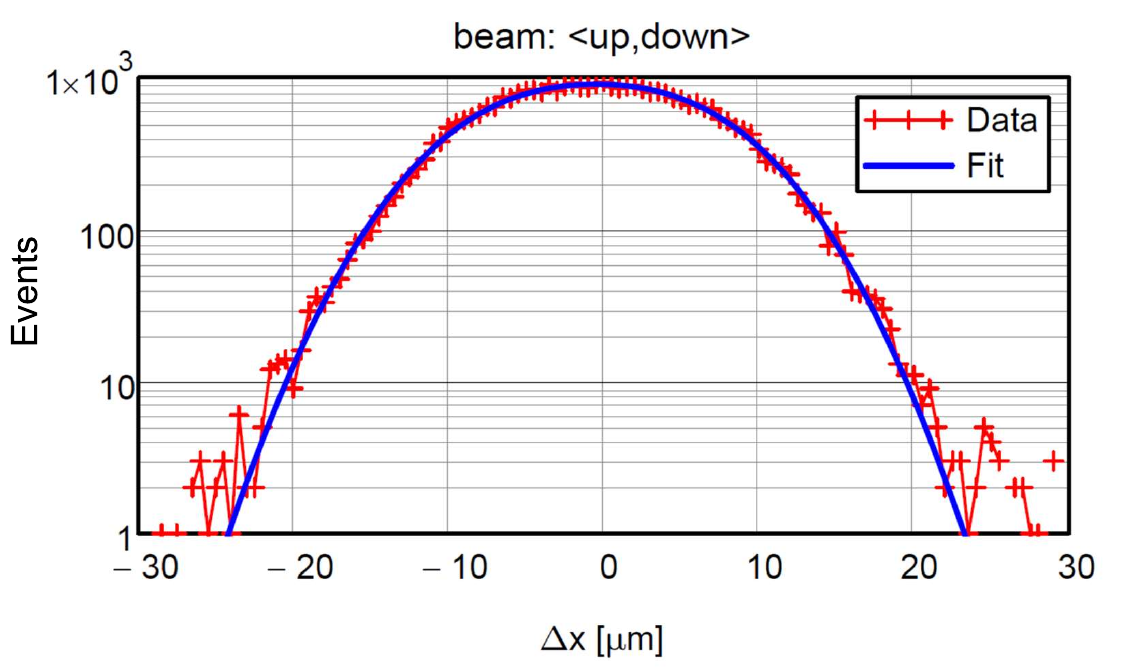}
    \label{fig:cls1xme}
   \end{subfigure}%
    ~
   \begin{subfigure}[a]{0.5\textwidth}
    \includegraphics[width=\textwidth]{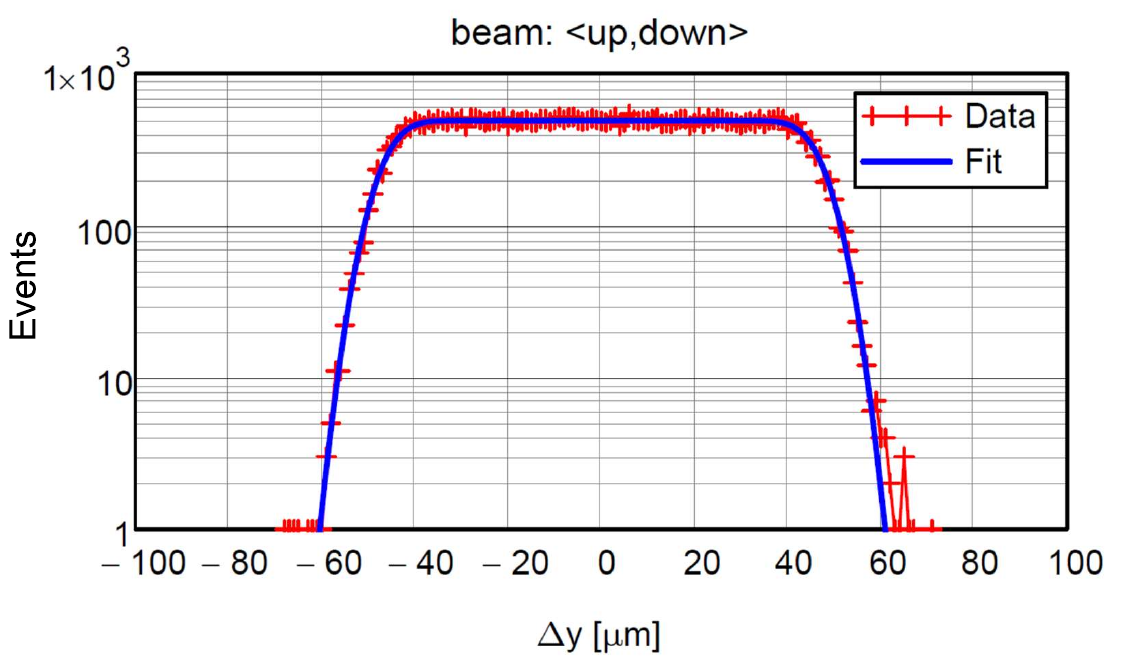}
    \label{fig:cls1yme}
   \end{subfigure}%

      \begin{subfigure}[a]{0.5\textwidth}
    \includegraphics[width=\textwidth]{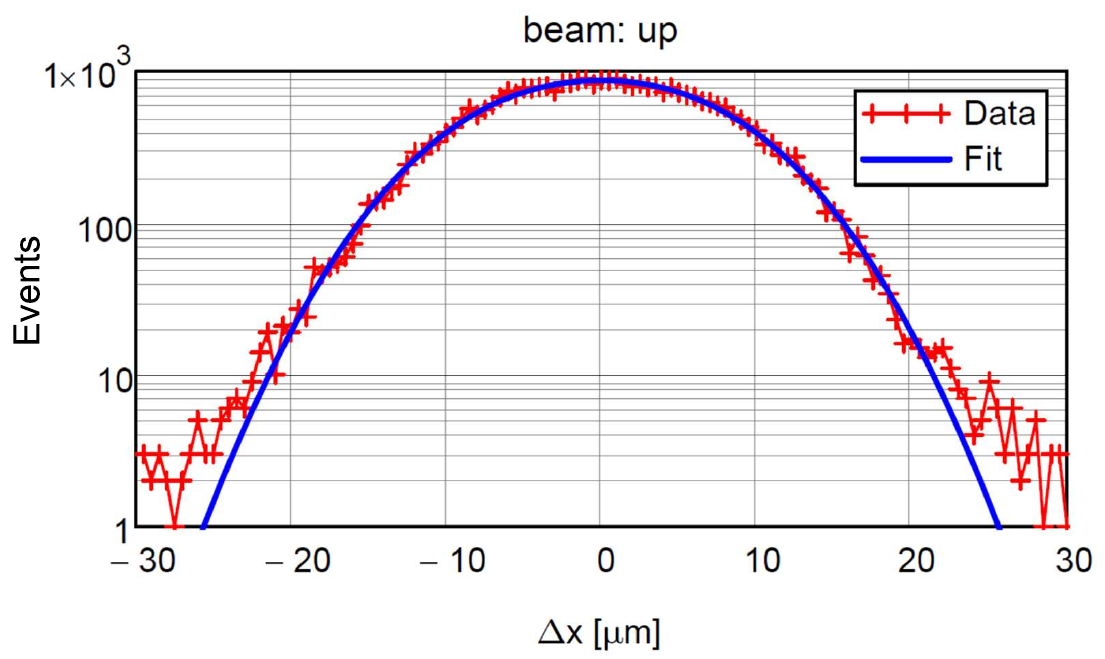}
    \label{fig:cls1xip}
   \end{subfigure}%
    ~
   \begin{subfigure}[a]{0.5\textwidth}
    \includegraphics[width=\textwidth]{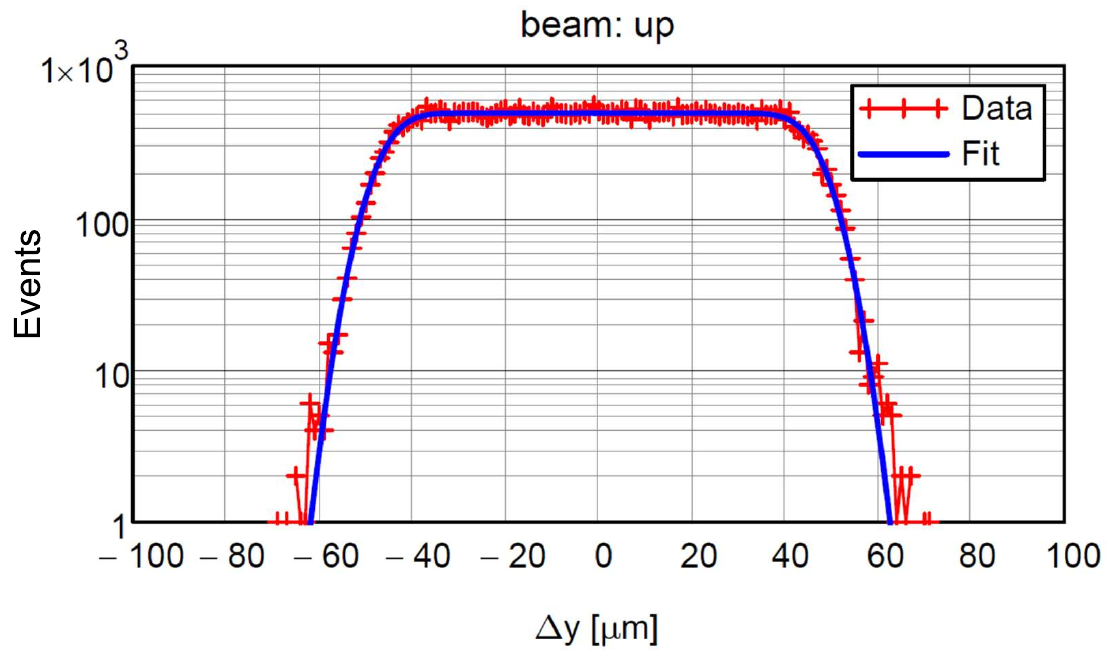}
    \label{fig:cls1yup}
   \end{subfigure}%

   \begin{subfigure}[a]{0.5\textwidth}
    \includegraphics[width=\textwidth]{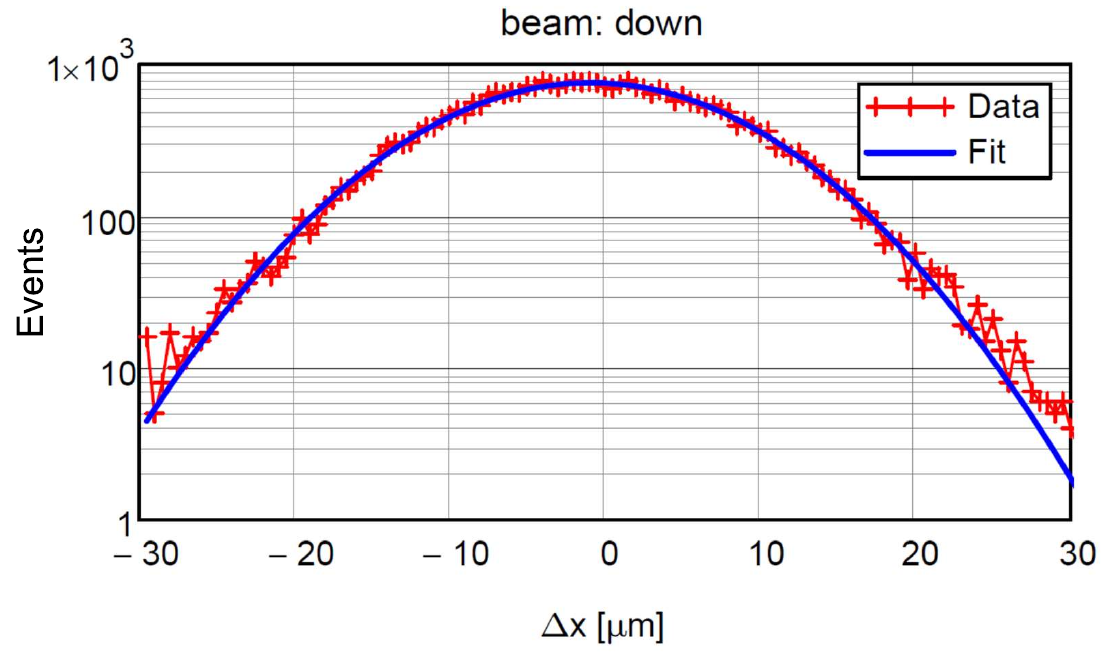}
    \label{fig:cls1xdo}
   \end{subfigure}%
    ~
   \begin{subfigure}[a]{0.5\textwidth}
    \includegraphics[width=\textwidth]{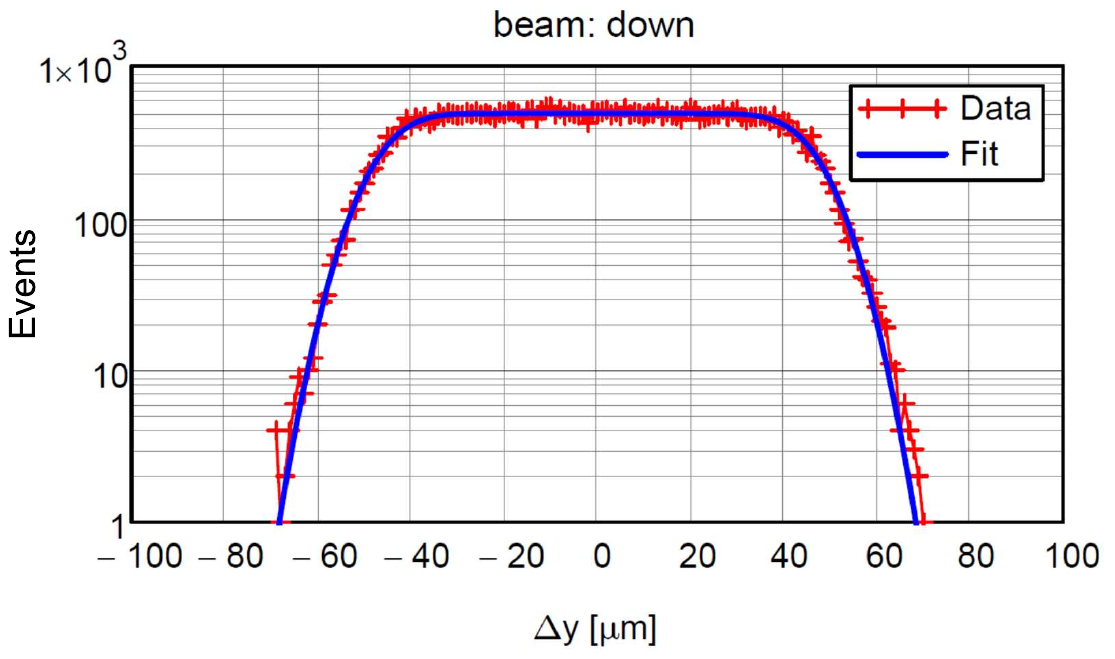}
    \label{fig:cls1ydo}
   \end{subfigure}%
   \caption{Residuals distributions $\Delta x = x_\mathrm{DUT} - x_\mathit{beam}$ (left) and $\Delta y = y_\mathrm{DUT} - y_\mathit{beam}$ (right), and the mean of the upstream and downstream (top), the upstream (middle) and the downstream (bottom) beam tracks at the DUT, for $\mathit{cls} = 1$ events.
   The continuous lines are the results of the fits of Eq.\,\ref{equ:erfs} to the data.
   The values of the parameters and their statistical uncertainties are shown in Table\,\ref{tab:Fitcls1}.}
  \label{fig:cls1xy}
 \end{figure}

Figure\,\ref{fig:cls1xy} shows for the $\mathit{cls} = 1$ events of the data of Section\,\ref{sect:Setup} the residual distributions of the position reconstructed in the DUT (pixel centres) and the beam track-position at the DUT for the \SI{25}{\um} ($x$) and the \SI{100}{\um} ($y$) directions, for the beam track-positions from the upstream and downstream beam telescope arms, and from their averages. Whereas in the $y$-direction the expected flat top is observed, it is absent in the $x$-direction. The reason is that for the \SI{100}{\um} pixels the root-mean-square of a flat distribution with the width of the pixel pitch, $\mathit{rms} = \mathit{pitch} / \sqrt{12} = \SI{28.9}{\um}$ is large compared to the track-position resolution of 4.5 to \SI{7.5}{\um}, which is not the case for the $x$-direction.

The continuous lines are fits by
 \begin{equation}\label{equ:erfs}
   f(x) = \frac{A}{2} \cdot \left( \mathrm{erf}\left(\frac {x - x_0 + w_x/2} {\sqrt{2} \, \sigma_x} \right) - \mathrm{erf} \left(\frac {x - x_0 - w_x/2} {\sqrt{2}\, \sigma_x} \right) \right)
 \end{equation}
to the residual distributions. The free parameters of the fits are the normalisation, $A$, the mean position and the full width of the box distribution, $x_0$ and $w_x$, and $\sigma_x$ the \emph{rms} of the convolution by the track-position resolution, which is assumed to be Gaussian. Similarly for $y$. The values of the fitted parameters are given in Table\,\ref{tab:Fitcls1}.

  \begin{table} [!ht]
   \caption{Values and statistical uncertainties of the parameters from fits of Eq.\,\ref{equ:erfs} to the data of Figure\,\ref{fig:cls1xy}.}
  \label{tab:Fitcls1}
  \centering
   \begin{tabular}{c||c|c|c||c|c|c}
   Beam & $x_0\,[\unit{\um}]$ & $w_x\,[\unit{\um}]$ & $\sigma_x\,[\unit{\um}]$
   & $y_0\,[\unit{\um}]$ & $w_y\,[\unit{\um}]$ & $\sigma_y\,[\unit{\um}]$ \\
  \hline
   $\langle \mathrm{up,down} \rangle$ & $0.47 \pm 0.04$ & $17.3 \pm 0.3$  & $4.88 \pm 0.12$
   & $-0.25 \pm 0.10$ & $94.5 \pm 0.2$  & $4.68 \pm 0.15$ \\

   up & $0.02 \pm 0.05$ & $17.1 \pm 0.5$  & $5.52 \pm 0.18$
      & $-0.37 \pm 0.12$ & $ 94.6 \pm 0.3$  & $5.20 \pm 0.18$ \\

   down & $ - $ & $ - $  & $ - $
   & $-0.04 \pm 0.16 $ & $ 94.6 \pm 0.3$  & $ 7.33 \pm 0.24$ \\
   \end{tabular}
 \end{table}

No parameter values are given for the fit for the downstream telescope arm in the $x$-direction. Although the fit converges and provides a valid description of the data, the error matrix is practically singular: The parameters $A$ and $w$ are correlated and the fitted values are not meaningful. This shows that the $\mathit{cls} = 1$\,method only works if the DUT spatial resolution $\sigma \lessapprox \mathit{pitch}/\sqrt{12}$. It is noted that the flat top is also absent for the other residuals in the $x$-direction, however the fits give reliable results. Comparing the results for $\sigma$ to the values of Table\,\ref{tab:Sig_non} of the $\mathit{cls} = 2$  analysis shows that the results agree within the statistical uncertainties; however, the values of the $\mathit{cls} = 1$ analysis are systematically higher by a fraction of a micrometer.

To conclude: for tracks at normal incidence, both $\mathit{cls} = 1$ and $\mathit{cls} = 2$ methods can determine precisely the position resolution of the tracks from beam telescope extrapolated to the DUT. However, the $\mathit{cls} = 1$ method cannot be used if the track-position resolution exceeds $ \approx \mathit{pitch}/\sqrt{12}$, and, in addition, a parametrization of the resolution function has to be assumed. The $\mathit{cls} = 2$ method does not have these limitations and determines the resolution function. So far, for the $\mathit{cls} = 1$ method a study of the influence of electronics noise, threshold, cross-talk, angular misalignment and energetic $\delta$-electrons has not been published. This paper shows for the $\mathit{cls} = 2$ method how these effects can be handled.

\end{document}